\documentclass[aps, pre, superscriptaddress, twocolumn]{revtex4-1}

\usepackage{amssymb, amsmath, amsfonts, amsthm, mathtools}
\usepackage{graphicx, xcolor}
\usepackage{empheq}
\usepackage[T1]{fontenc}
\usepackage{xspace}
\usepackage{epstopdf}

\usepackage{soul}

\usepackage{tikz}
\usetikzlibrary{calc}
\usepackage{marvosym}
\usepackage{textcomp}
\usepackage{color}

\renewcommand*{\d}{{\mathrm d}}

\newcommand*{\xth}{x_\mathrm{th}\xspace}
\newcommand*{\xr}{x_\mathrm{r}\xspace}

\newcommand{\eq}[1]{Eq.~#1}
\newcommand{\eqs}[1]{Eqs.~#1}
\newcommand{\fig}[1]{Fig.~#1}
\newcommand{\figs}[1]{Figs.~#1}

\renewcommand{\sec}[1]{Sec.~#1}

\begin{document}

\title{Interspike interval correlations in networks of inhibitory integrate-and-fire neurons}

\author{Wilhelm Braun}
\email{wilhelm.braun@cantab.net}
\affiliation{Institut f\"{u}r Genetik, Universit\"{a}t Bonn, Kirschallee 1, 53115 Bonn, Germany}
\affiliation{Department of Physics and Centre for Neural Dynamics, University of Ottawa, 598 King Edward, Ottawa K1N 6N5, Canada}
\author{Andr\'{e} Longtin}
\affiliation{Department of Physics and Centre for Neural Dynamics, University of Ottawa, 598 King Edward, Ottawa K1N 6N5, Canada}

\date{\today}

\begin{abstract}
We study temporal correlations of interspike intervals (ISIs), quantified by the network-averaged serial correlation coefficient (SCC), in networks of both current- and conductance-based purely inhibitory integrate-and-fire neurons. Numerical simulations reveal transitions to negative SCCs at intermediate values of bias current drive and network size. As bias drive and network size are increased past these values, the SCC returns to zero. The SCC is maximally negative at an intermediate value of the network oscillation strength. The dependence of the SCC on two canonical schemes for synaptic connectivity is studied, and it is shown that the results occur robustly in both schemes. For conductance-based synapses, the SCC becomes negative at the onset of both a fast and slow coherent network oscillation. Finally, we devise a noise-reduced diffusion approximation for current-based networks that accounts for the observed temporal correlation transitions.
\end{abstract}


\maketitle


\section{Introduction}
Quantifying the statistics of spiking in single neuron models is an important goal in computational neuroscience. A large body of work has dealt with first-order statistics of interspike intervals in canonical single neuron models, often computed in the context of memory-free renewal first-passage problems \cite{gerstein_mandelbrot_1964}.
In recent years, progress has been made in tackling the much harder problem of second-order spiking statistics in non-renewal neuron models, among them the serial correlation coefficient (SCC) at lag $k$, which is defined as the Pearson correlation between ordered ISI sequences shifted by an amount $k$, e.g., for $k=1$, the SCC at lag $1$ is the correlation coefficient between adjacent ISIs. Its computation has uncovered enhanced information transmission in single neurons via the effect of noise shaping \cite{chacron_lindner_longtin_2004, *avila_akerberg_chacron_2011, *nesse_maler_longtin_2010, *chacron_longtin_maler} and increased detectability of weak signals \cite{ratnam_nelson_2000}. In particular, it is now possible to compute, in some cases even analytically, the SCC for single-neuron models in the presence of spike-frequency adaptation \cite{urdapilleta_pre_onset_of_correlations_2011, *schwalger_lindner_2013, *urdapilleta_2016}, colored noise \cite{schwalger_droste_lindner_2015, *mankin_rekker_2016} and time-dependent deterministic or stochastic firing thresholds \cite{benda_maler_longtin_2010, *braun_matthews_thul_2015, *urdapilleta_pre_2011, *urdapilleta_jphysa_2012}. Recently, it was also shown that, at short observation times, negative ISI correlations can enhance the resolution of a nonlinear dynamical sensor whose design was inspired by a simple non-renewal neuron model \cite{nikitin_stocks_bulsara_2012}. There also have been recent efforts to characterize the patterning of ISI sequences using ordinal analysis, revealing parameter sets that maximize the probability of certain patterns \cite{reinoso_torrent_masoller_2016, *masoliver_masoller_2018}.

In contrast, very few studies deal with the computation of temporal correlations in networks of neurons. In \cite{muller_et_al_2007}, a theory to compute the SCC in networks of adapting neurons based on hazard functions is put forward; the main focus of that study is the computation of first-order statistics, such as the mean ISI or the average activity of neurons in the network. When it comes to the computation of SCCs in networks, a notable exception are the recent studies \cite{dummer_wieland_lindner_2014, *pena_et_al_2018}, were the main focus is on the self-consistent computation of the spectra of neuronal spike trains in asynchronous networks of excitatory and inhibitory neurons. Iterative numerical methods are described to compute the spike train power spectrum self-consistently simulating one single representative neuron only. This adds to previous analyses of network synchrony in terms of a single ``effective'' neuron \cite{brunel_hakim_1999}. Because the approach directly simulates spike trains, the SCC is also computed and shows weak positive and negative values. However, the validity of the proposed scheme is limited to the asynchronous state without a global oscillation of the network activity. Moreover, there are a number of studies dealing with properties of networks in which each constituent neuron is endowed with an adaptation mechanism as an intrinsic correlation-generating, highlighting the benefits of adaptation for information transmission, reliable neural coding and noise shaping \cite{farkhooi_muller_nawrot_2011, *farkhooi_et_al_2013, mar_et_al_1999}.

Here we focus solely on the ISI correlations of single neurons in networks that transition from asynchronous to synchronous behavior. The motivation to study this setup is twofold. First of all, it is desirable to understand how the statistics of a neuron embedded in a network changes compared to the well-studied isolated case, because neuronal coding often relies on populations of neurons. In particular, in networks of neurons without correlation-generating intrinsic mechanisms, under what conditions do correlations arise and how are they maximized? Secondly, precise spiking patterns generated by neurons embedded in a network can potentially be selected by post-synaptic plasticity mechanisms in target cells downstream from the network, e.g. spike-timing dependent plasticity \cite{bi_poo_1998, sjostrom_et_al_2001, kempter_gerstner_van_hemmen_1999} or short-term facilitation and depression \cite{abbott_regehr_2004, tsodyks_markram_1997}, leading to enhanced weak signal detectability \cite{luedtke_nelson_2006}. It is thus desirable to understand how networks of spiking neurons generate temporal correlations in the constituting neurons' output spike trains.
Our study explores the array of questions above by focusing solely on very short range ISI correlations (in fact, only correlations between neighboring ISIs), for the sake of conciseness. We stress from the outset that in strongly oscillatory network states, correlations extend beyond more than one lag, as e.g. observed in \cite{neiman_russell_2005, bauermeister_schwalger_et_al_2013}. Also, we do not consider correlations between neurons in this study, which is set in the context of sparse networks. A large body of work exists that describes the pairwise correlation coefficient between the spike trains or membrane voltages of two different neurons in networks with different topologies \cite{ginzburg_sompolinsky_1994, moreno_bote_parga_2006, tetzlaff_et_al_2008, helias_tetzlaff_diesmann_2014}.

The paper is organized as follows. In \sec \ref{sec:model}, we present the network model, synaptic connectivity rules and the central quantities of interest for this study, among them the mean network SCC. We describe the parameter dependence of the mean network SCC at the onset of a coherent global network oscillation in \sec \ref{sec:mean_network_scc}. Whereas the focus in this section is on networks with current-based synapses, we briefly extend our findings to networks with conductance-based synapses that are capable of generating both slow and fast gamma oscillations. Finally, in \sec \ref{sec:DA}, we explore how the mean network SCC can be computed by an effective single offline neuron receiving time-dependent input recorded online during a full network simulation. We conclude with a brief discussion of our results in \sec \ref{sec:discussion}.

\section{Network of inhibitory neurons}
\label{sec:model}

\subsection{Neuron and synaptic dynamics}

\subsubsection{Networks with current-based synapses}

We consider $N$ leaky integrate-and-fire (LIF) neurons with membrane voltages $X_{i}$, $i \in [1,2,..., N]$, with $X_{i} \in (-\infty, \xth ]$. They evolve according to \cite{brunel_hakim_1999}

\begin{equation}
 \frac{\d X_{i}}{\d t} = \gamma_{t}\left[I_{0} - X_{i} + I_{i}^{\mathrm{syn}}(t)\right] + \sqrt{\gamma_{t}} \sigma_{0} \xi_{i}(t) \, .
 \label{eq:definition_X}
\end{equation}

For conciseness, a factor carrying units of resistance in front of $I_{0}$ and $I_{i}^{\mathrm{syn}}$ has been set to one and omitted. When $X_{i}(t)$ reaches the threshold $\xth$ from below, a spike is recorded and $X_{i}$ is reset to $\xr$ immediately; when we consider a refractory period of length $\tau_{r}$, the neuron is kept at this membrane voltage for a time $\tau_{r}$. The membrane time constant is $\tau_{t} = \frac{1}{\gamma_{t}}$. $\xi_{i}$ is a standard zero-mean Gaussian white noise. If not mentioned otherwise, we choose $\xr = 10~\text{mV},~ \xth = 20~\text{mV},~ \tau_{t} = 20~\text{ms}$.

The current-based inhibitory synaptic current is given by
 
\begin{equation}
I_{i}^{\mathrm{syn}}(t) = \tau_{t} J \sum_{k = 1}^{C} \sum_{j}\delta(t - t^{k}_{j} -D) \, .
\label{eq:synaptic_current}
\end{equation}

$J<0$ is the \textit{inhibitory} synaptic strength. We choose $J = -0.1~\text{mV}$ and fix the synaptic delay $D = 2~\text{ms}$. Hence, the outer sum in \eq \ref{eq:synaptic_current} is over $C$ neurons in the presynaptic neighborhood of size $C$, while the inner sum is over the spikes $j$ of the neuron $k$ at times $t^{k}_{j}$. For more details on implementation, we refer to Appendix \ref{sec:simulation_details}.

Finally, $I_{0}$ is the external bias current, and $\sigma_{0}$ is the strength of the external noise. If not mentioned otherwise, we choose $\sigma_{0} = 1~\text{mV}$. The single neuron fires periodically in the absence of private noise and synaptic input (i.e. is in the {\it suprathreshold regime}) if $I_{0} > 20~\text{mV}$.

In the diffusion approximation, we may express the synaptic current as follows \cite{brunel_hakim_1999}:

\begin{equation}
I_{i}^{\mathrm{syn}}(t) = \mu(t) + \sigma(t) \sqrt{\tau_{t}} \xi_{i}(t) \, ,  
\label{eq:diffusion_approximation}
\end{equation}

with 

\begin{equation}
 \mu(t) = \mu_{\text{loc}}\, ,
 \label{eq:drift_term}
\end{equation}

where the local part of the average synaptic current is related to the firing rate $\nu$ of the network at time $t-D$:  $\mu_{\text{loc}} = C J \nu (t-D) \tau_{t} $. The strength of the fluctuations is similarly given by

\begin{equation}
\sigma(t) = \sigma_{\text{loc}}\, , 
\label{eq:diffusion_term}
\end{equation}

with $\sigma_{\text{loc}} = |J| \sqrt{C \nu(t-D) \tau_{t}}$. 

\subsubsection{Networks with conductance-based synapses}
For networks with conductance-based synapses, the membrane voltage evolves according to

\begin{equation}
C_{I}\frac{\d X_{i}(t)}{\d t}= g^{I}_{l}\left(E^{I}_{\text{rest}} - X_{i}(t)\right) + g^{I}_{\text{inh}}(t)(E^{I}_{\text{inh}} -  X_{i}(t)) + I_{0} \, .
\end{equation}

Every neuron receives inputs from other inhibitory neurons in the network via synapses with time-dependent conductance $g^{I}_{\text{inh}}(t)$, whose time course is given by a bi-exponential function:

\begin{equation}
g^{I}_{\text{inh}}(t) = g^{I}_{\text{inh}, \text{peak}}s^{I}_{\text{inh}} \left[\exp\left(-\frac{t-\tau_{l}}{\tau^{I}_{\text{inh}, d}} \right) - \exp \left(-\frac{t-\tau_{l}}{\tau^{I}_{\text{inh}, r}} \right) \right] \, ,
\label{eq:ginh}
\end{equation}

for $t \geq \tau_{l}$, $g^{I}_{\text{inh}}(t) = 0$ otherwise. Here, $s^{I}_{\text{inh}}$ is a constant that ensures $g^{I}_{\text{inh}}$ reaches its maximum $g^{I}_{\text{inh}, \text{peak}}$ (which is defined by computing the maximum of the term in brackets in \eq \ref{eq:ginh}):

\begin{equation*}
 \frac{1}{s} = \left( \frac{\tau_{r}}{\tau_{d}}\right)^{\frac{\tau_{r}}{\tau_{d}-\tau_{r}}} - \left( \frac{\tau_{r}}{\tau_{d}} \right)^{\frac{\tau_{d}}{\tau_{d}-\tau_{r}}} \, ,
\end{equation*}

where we set $\tau_{r} \equiv \tau^{I}_{\text{inh}, r}$, $\tau_{d} \equiv \tau^{I}_{\text{inh}, d}$ and $s \equiv s^{I}_{\text{inh}}$. With these definitions, increasing $\tau^{I}_{\text{inh}, r}$ has the effect of prolonging the time $g^{I}_{\text{inh}}(t)$ needs to reach its maximum, and therefore extends the effect of one pre-synaptic inhibitory pulse by increasing the area under the time course of $g^{I}_{\text{inh}}(t)$. For all remaining parameter values, we refer to Appendix \ref{sec:coba}.

\subsection{Synaptic connection rules}
There are different strategies to choose the number of presynaptic neurons $C$ (see \eq \ref{eq:synaptic_current}) per neuron. When we only fix the connection probability $p$, not all neurons have the same number of inhibitory synapses. Excluding autapses, the distribution of the number of synapses onto a neuron (i.e. the in-degree) follows a Bernoulli distribution $\mathcal B(N-1, p)$ with mean $\overline{C} = p(N-1)$ and standard deviation $\sqrt{(N-1) p (1-p)}$. We will call this scenario the $P$-fixed case. For $N = 1000$ and $p = 0.2$, the standard deviation is $\approx 13 $, whereas the mean is $ \overline{C} \approx 200$. For more details on the implementation, we refer to Appendix \ref{sec:connectivity_details}.

This scenario is in contrast to the case where every neuron has exactly $C$ inhibitory synapses. This is the $C$-fixed case. We always choose $C = \overline{C}$ for comparisons of the $P$-fixed vs $C$-fixed connectivity scenarios. We will see below that these two different prescriptions for synaptic connectivity have a profound influence on the second-order statistics of the network, as e.g. reflected in the mean and standard deviation of the distribution of the SCC across neurons, which we now define.

\subsection{The mean network SCC and ISI}
For a single spike train of neuron $i$, the SCC $\rho$ is defined as


\begin{equation}
\rho^{i}(n, k) = \frac{\mathbb E \left( T^{i}_{n} T^{i}_{n+k}\right) -\mathcal Q^{i}_{1}(n,k)}{\mathcal Q^{i}_{2}(n,k)} \, ,
\label{eq:definition_SCC}
\end{equation}

where

\begin{equation*}
\mathcal Q^{i}_{1}(n,k) = \mathbb E (T^{i}_{n}) \mathbb E(T^{i}_{n+k}) \, ,
\end{equation*}

and
\begin{equation*}
\mathcal Q^{i}_{2}(n,k) = \sqrt{\mathrm{Var}(T^{i}_{n})\mathrm{Var}(T^{i}_{n+k})} \, .
\end{equation*}

Here, $T^{i}_{n}$ denotes the $n$th ISI in the spike train of neuron $i$. Furthermore, $\mathbb E(...)$ denotes the expectation across time for one single neuron. In general, the SCC depends on both the lag $k$ and the position $n$ of the ISI in the spike train \cite{mankin_rekker_2016, braun_thul_longtin_2017}. We here consider only stationary spike trains, for which the SCC depends only on the lag $k$ between different ISIs. We will denote this quantity by $\rho(k)$  and focus on $k = 1$, i.e. on successive ISI correlations. Significant short-term correlations between ISIs will be typically reflected in deviations of this coefficient from zero.
The main quantity of interest in this paper, which we call the \textit{mean network SCC}, is the neuron-averaged SCC in a network of $N$ neurons:
%

\begin{equation}
\rho(k=1) = \frac{1}{N} \sum_{i = 1}^{N} \rho^{i}(k=1) \, .
\label{eq:network_scc}
\end{equation}

The standard deviation of the SCC across neurons is defined as 
%

\begin{equation}
\text{std}\left(\rho\right) = \sqrt{\frac{1}{N}\sum_{i = 1}^{N} \left( \rho^{i}(k=1) - \rho(k=1)   \right)^{2}} \, .
\label{eq:network_scc_std}
\end{equation}

It is used in the following to quantify the spread of SCC values at lag $1$ across the population.

The second quantity of interest is the mean network ISI, which is defined as the mean of the average ISI across neurons:


\begin{equation}
\langle T \rangle = \frac{1}{N} \sum_{i=1}^{N} \mathbb E(T^{i}) \, ,
 \label{eq:network_ISI}
\end{equation}

where  $T^{i}$ is the set of stationary ISIs for neuron $i$. Again, the standard deviation of the mean ISI, $\text{std}(\langle T \rangle)$, is defined analogously. 

\subsection{Population activity and power spectra}
We denote the population activity $\sum_{i = 1}^{N}\sum_{j}\delta(t-t^{i}_{j})$ of a network of purely inhibitory neurons by $\nu(t)$. 
The power spectral density (PSD) of the population activity is computed from the population activity $\nu(t)$. Similarly, the averaged single-neuron spectrum $\langle S_{xx}(f) \rangle$ is computed from all spikes of the neuron group. It is the average PSD of spike trains $x_{i} = \sum_{j}\delta(t-t^{i}_{j})$, scaled such that as $f \rightarrow \infty$, $\langle S_{xx}(f) \rangle$ approaches the inverse mean ISI given by \eq \ref{eq:network_ISI}. For more details on the implementation, we refer to Appendix \ref{sec:spectral_measures}. Formal definitions of the population PSD and the averaged single neuron power spectrum can be found e.g. in \cite{lindner_schimansky_longtin_lif}. 

\section{Behavior of the mean network SCC with varying network size and bias current}
\label{sec:mean_network_scc}
In the presence of a noisy network oscillation, one expects the SCC of a single neuron in the network to be negative. A simple toy problem illustrates this point. Assume that firing times $t_{i}$ are generated according to the following prescription: $t_{i} = i \langle t\rangle + \sigma \xi(i)$, where $\langle t \rangle$ is the period of the oscillation and $\sigma>0$ is a parameter that determines how strong the independent standard normal Gaussian random variables $\xi(i)$ influence the dynamics. We show in Appendix \ref{sec:appendix_1} that the SCC between adjacent ISIs in this case is negative: $\rho(k=1) =  -\frac{1}{2}$. To explain this result, note that when an ISI larger than the mean $\langle t\rangle$ is generated, we must have a large value of $\xi$. Because the random variables $\xi(i)$ are independent, it is unlikely that the next firing time $t_{i+1}$ will be larger than the mean again; instead, it will more likely lie around the mean, therefore introducing negative correlations between adjacent ISIs in the spike train. In real networks of spiking neurons, the periodic oscillation will furthermore modulate the membrane potential, and thus spiking will occur preferentially near the phases corresponding to the maximal values of the network oscillation. 

A different mechanism occurs in sparse networks of LIF neurons. In these networks, a global oscillation of the network activity $A(t)$ develops as the bias current $I_{0}$ is increased \cite{brunel_hakim_1999}. The frequency of this oscillation is approximately independent of $I_{0}$ and depends mainly on the synaptic delay $D$ \cite{brunel_hakim_1999}.

\begin{figure}[h!]
\includegraphics[width = 0.45\textwidth]{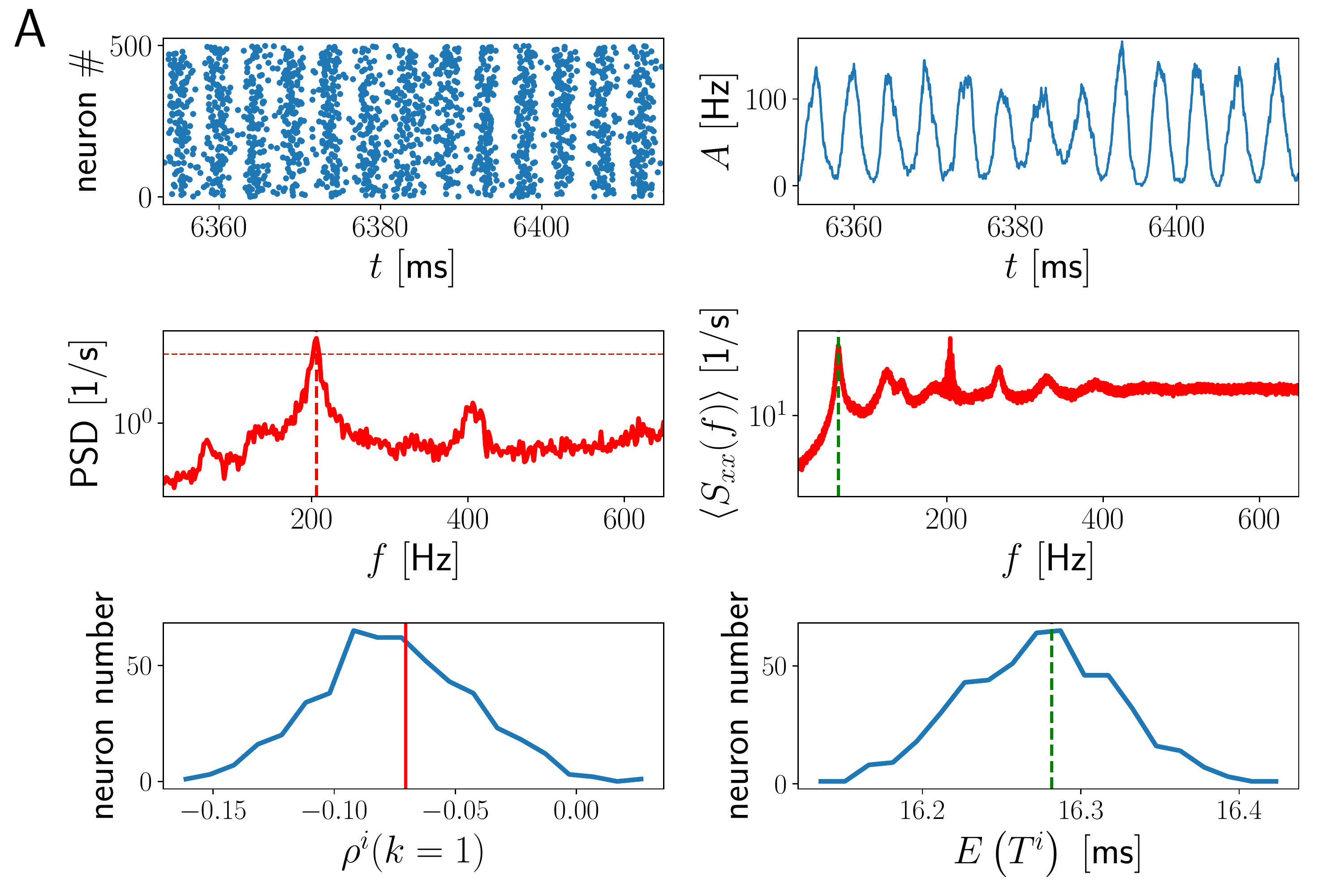}
\includegraphics[width = 0.45\textwidth]{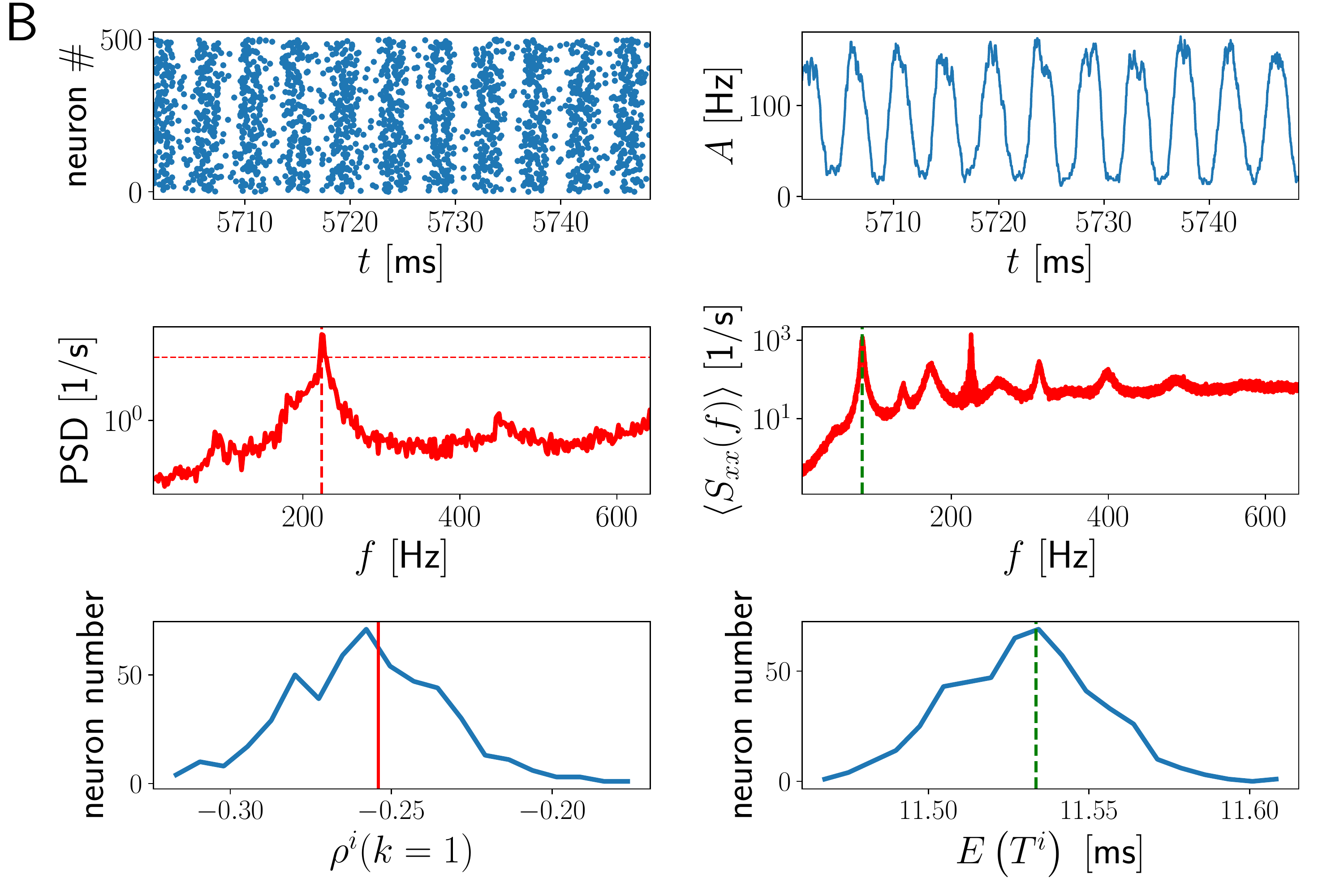}
\includegraphics[width = 0.45\textwidth]{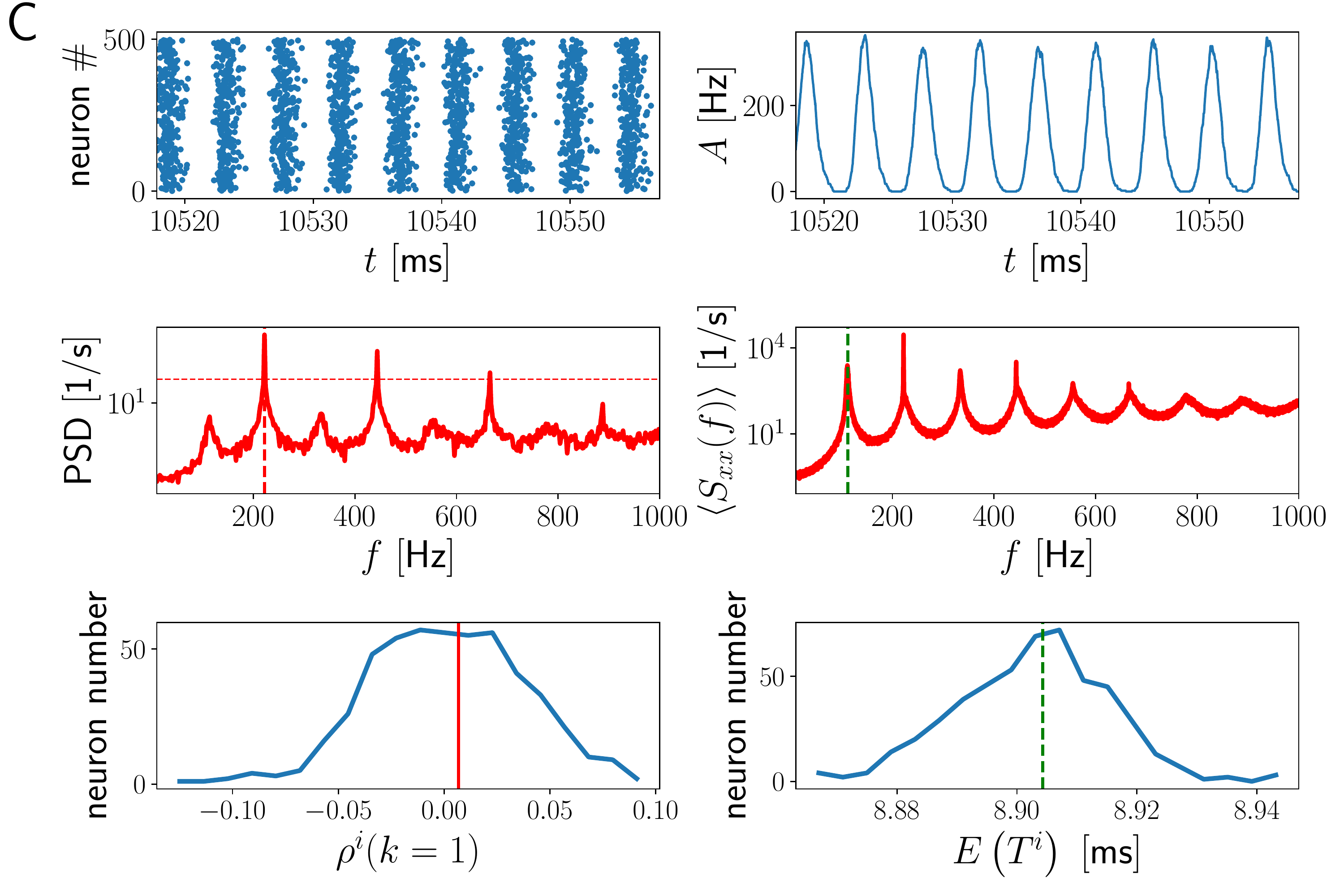}
\caption{Activity in networks with current-based synapses and SCC distributions at the onset of negative mean network SCCs.
Network activity (spike rastergram, network activity $A(t)$, power spectral density of network activity and average single-neuron spectrum $\langle S_{xx}(f) \rangle$), SCC distribution and mean ISI distribution for $N = 500$ for three different values of $I_{0}$: panels \textbf{A, B, C}: $I_{0} = 40,~50,~60~\text{mV}$ respectively, which corresponds to the location of the three white stars in \fig \ref{fig_2}. The mean network SCC is maximally negative at $I_{0} = 50~\text{mV}$, i.e. when power at the network frequency (horizontal red dashed lines in plots for PSD are at a fixed value of $10^{2}$ for comparison) increases. Remaining parameter values are as in \fig \ref{fig_2}. The green dashed lines show the mean network ISI (or its inverse in the average single-neuron power spectrum). No refractory period.}
\label{fig_1}
\end{figure}

In \fig \ref{fig_1}, we show that the mean network SCC turns negative at the onset of a coherent network oscillation with frequency $f \approx 200~\text{Hz}$ (vertical red dashed line in plots for the PSD in \fig \ref{fig_1} A-C); parameters can be adjusted to produce much slower rhythms as well. The average single-neuron spectrum $\langle S_{xx}(f) \rangle$ has a larger peak (green dashed line) at the inverse of the mean network ISI (approximately $61~\text{Hz}$), and another smaller peak at the network frequency $f$. This entails that an individual neuron in the network does not participate in every up-state of the network- instead, it skips cycles of the global oscillation. As the network oscillates more synchronously with increasing $I_{0}$, the first peak in the average single-neuron power spectrum decreases.

A state of high synchrony, in which each neuron has a higher probability of firing per cycle, i.e. skipping is nearly absent, is seen for larger $I_{0}$ (\fig \ref{fig_1} C) and the mean network SCC moves back towards zero. Hence we see a non-monotonic change in the mean network SCC as a function of $I_{0}$. At the onset of negative mean network SCCs (\fig \ref{fig_1} C), we have verified that the neuron-averaged SCC at lag $2$ (\eq \ref{eq:definition_SCC} for $k=2$) is slightly positive ($ \approx 0.07$) and at lag $3$, it is slightly negative ($\approx-0.017$), which is consistent with the underlying network oscillation causing the negative mean network SCC.

\subsection{SCC in networks with current-based synapses}
We restrict our numerical simulations to the suprathreshold regime and intermediate values of $I_{0}$ and $N$. Increasing network size or bias current past these values can result, respectively, in unrealistically synchronous network states (for which the SCC is either zero or positive) or strong inhibition that drives neurons far below their reset potential $\xr$, which is biologically not plausible.

\subsubsection{Onset of negative mean network SCCs}
\begin{figure}[h!]
\includegraphics[width = 0.5\textwidth]{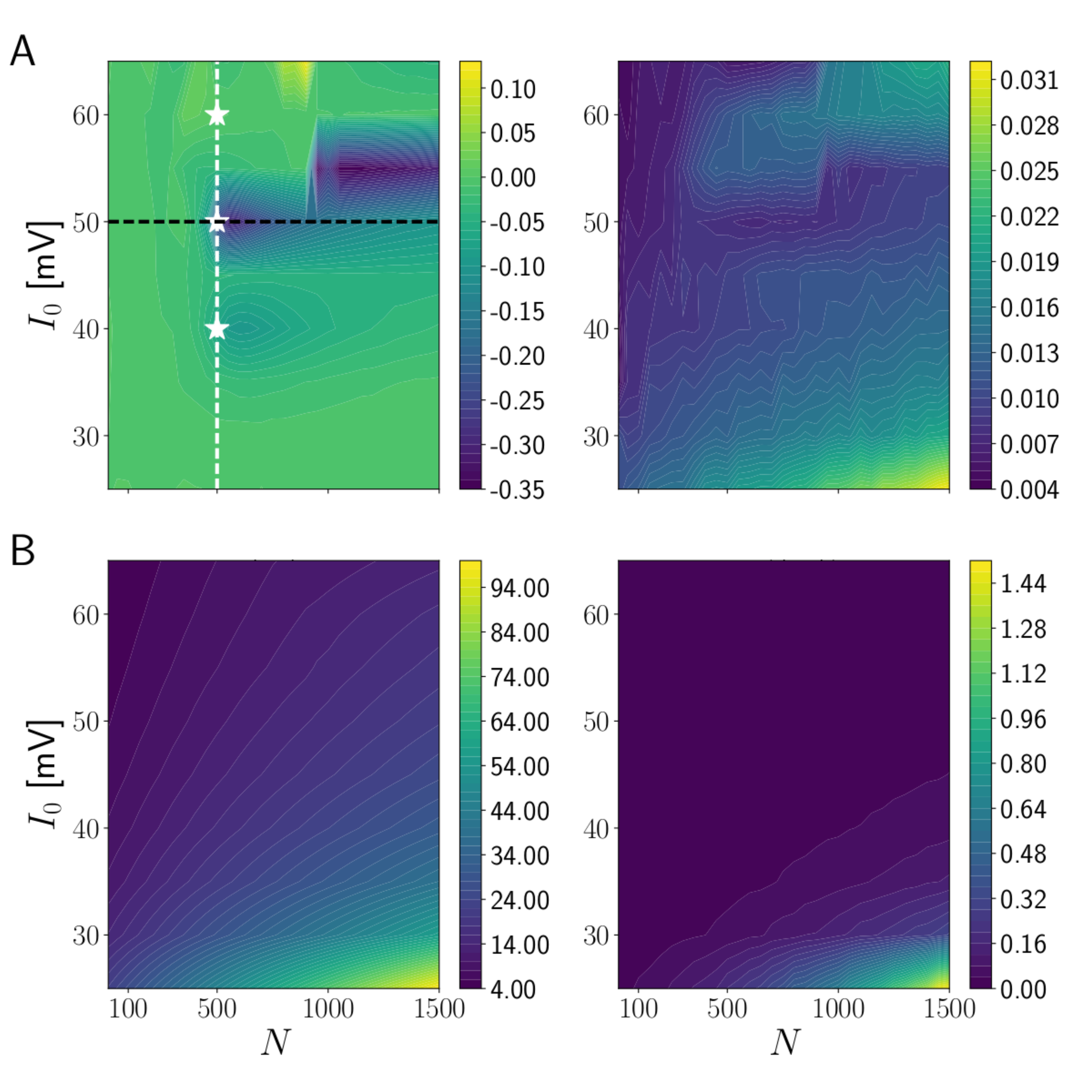}
\caption{Temporal correlation-decorrelation transitions as a function of network size $N$ and external bias $I_{0}$ for the $C$-fixed scenario in networks with current-based synapses. Mean network SCC (\eq \ref{eq:network_scc}) (\textbf{A}, left) and ISI (\eq \ref{eq:network_ISI}) (\textbf{B}, left) together with standard deviations of SCC (A, right) and of the mean ISI distribution across neurons (B, right) in the $C$-fixed connectivity scenario. The number of synapses onto each neuron is fixed at $C=p (N-1)$ with $p=0.2$. $d = 200~\text{s}$ simulation time. No refractory period.}
\label{fig_2}
\end{figure}

\begin{figure}[!h]
\includegraphics[width = 0.5\textwidth]{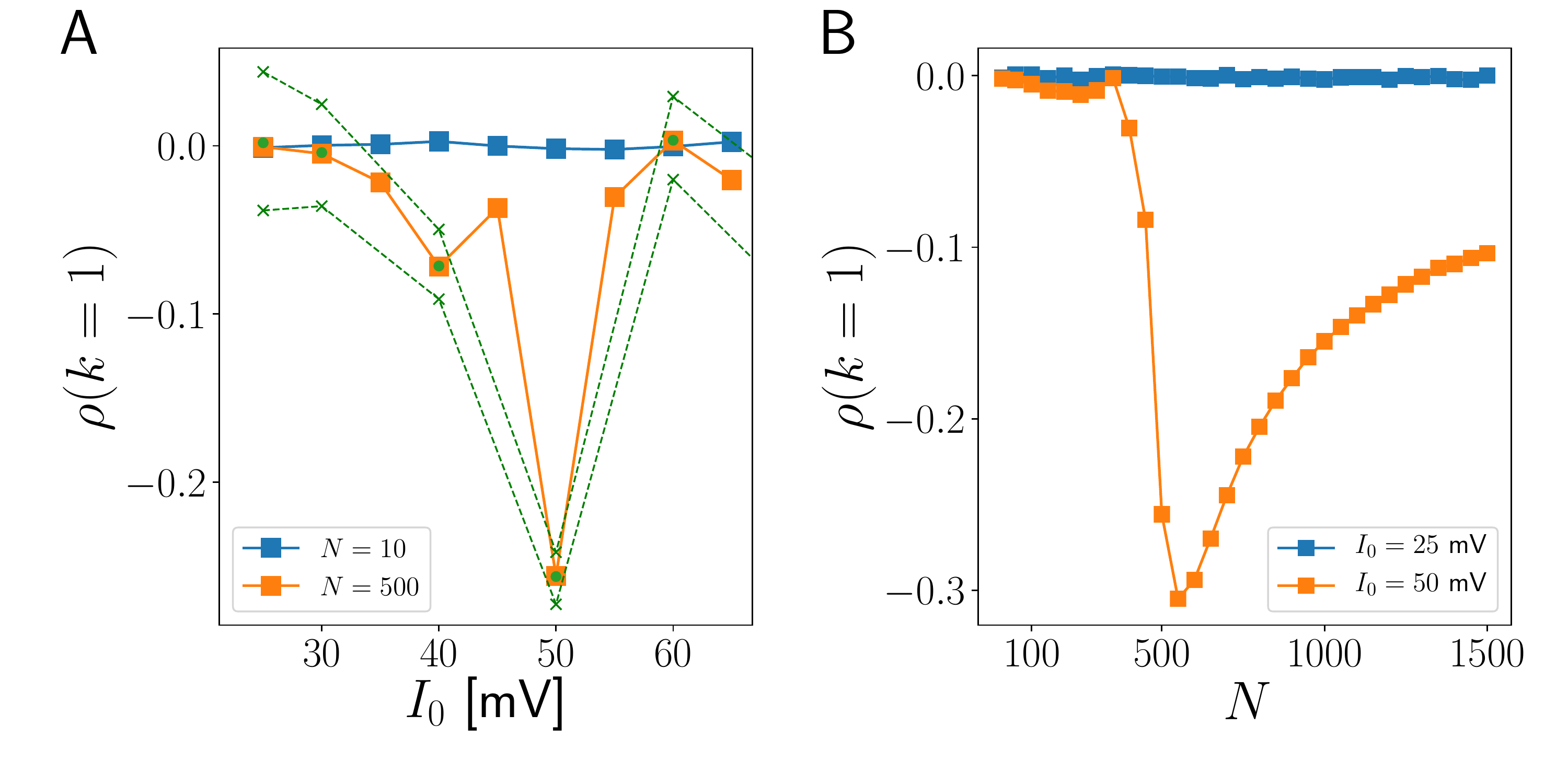}
\caption{Two types of temporal correlation-decorrelation transitions are present for the $C$-fixed case in networks with current-based synapses. Lineout of \fig \ref{fig_2} A along the white dashed vertical line (\textbf{A}, orange squares) and along the black dashed horizontal line (\textbf{B}, orange squares). Small green circles in A are for a control simulation with a different random seed and hence for a different network connectivity. Also, the duration of the control simulation was increased to $d = 300~\text{s}$ in contrast to \fig \ref{fig_2}, where $d = 200~\text{s}$. The green dashed lines are the minimum and the maximum of the SCC distribution for the control simulation. The blue squares are for smaller values of $N=10$ (panel A) or $I_{0}=25~\text{mV}$ (panel B) before the transition to negative mean network SCCs.}
\label{fig_3}
\end{figure}

In \fig \ref{fig_2}, we show the dependence of the mean network SCC and ISI on network size $N$ and bias current drive $I_{0}$. As $N$ increases for an intermediate value of $I_{0}$, the mean network SCC goes from around zero to negative values and then back to zero (\fig \ref{fig_2} A). In contrast, the mean network ISI (\fig \ref{fig_2} B) increases (decreases) monotonically with $N$ ($I_{0}$). Standard deviations of the two quantities across neurons (\eq \ref{eq:network_scc_std}) remain small (\fig \ref{fig_2}A, B right). Lineouts of \fig \ref{fig_2} along the white and black dashed lines, together with control simulations with a different detailed connectivity, are shown in \fig \ref{fig_3}, where it is also shown that the mean network SCC behaves non-monotonically as a function of $I_{0}$, too (cf. \fig \ref{fig_1}).

In summary, we observed a transition to negative mean network SCCs as the bias drive $I_{0}$ is increased. 

\begin{figure}[h!]
\includegraphics[width = 0.4\textwidth]{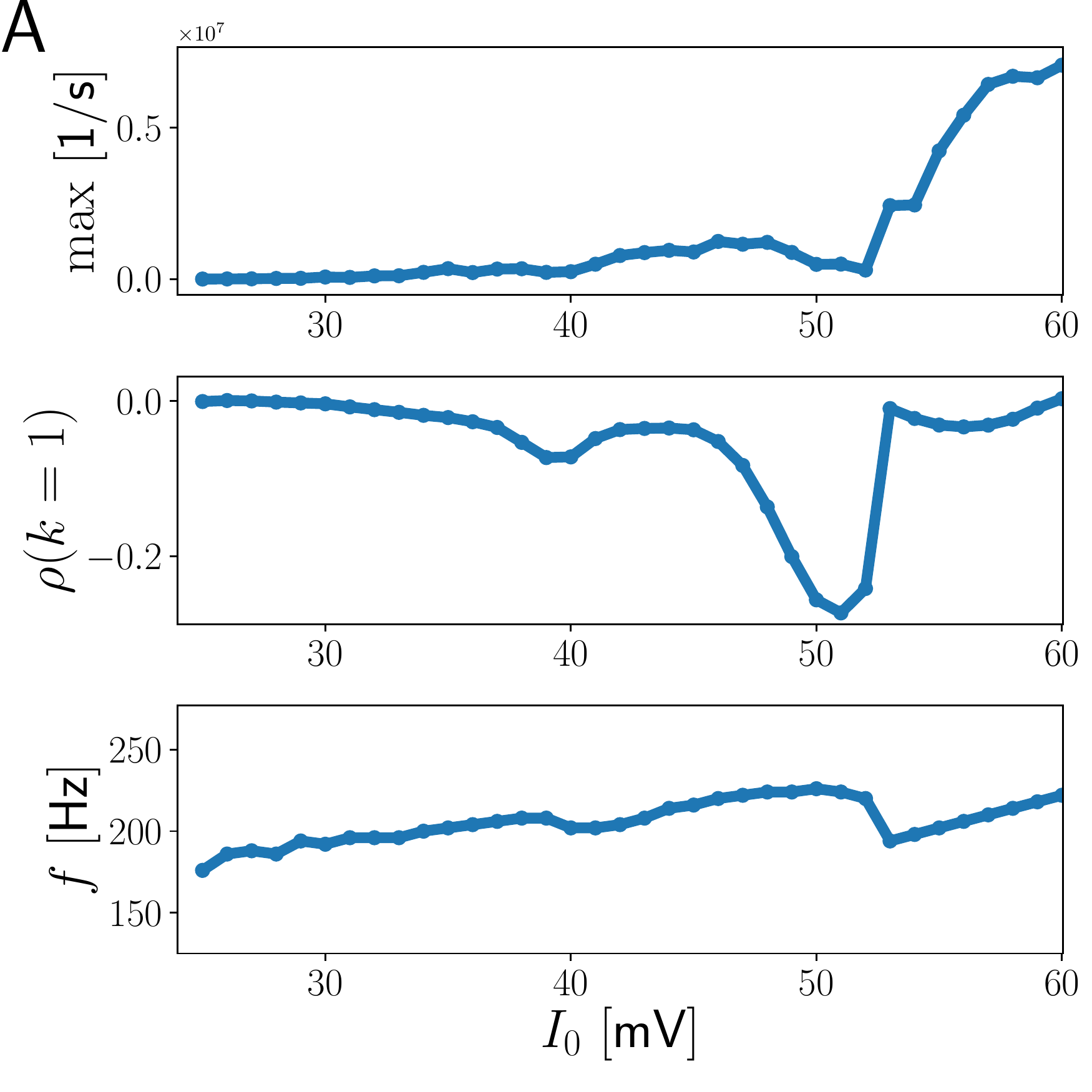}
\includegraphics[width = 0.4\textwidth]{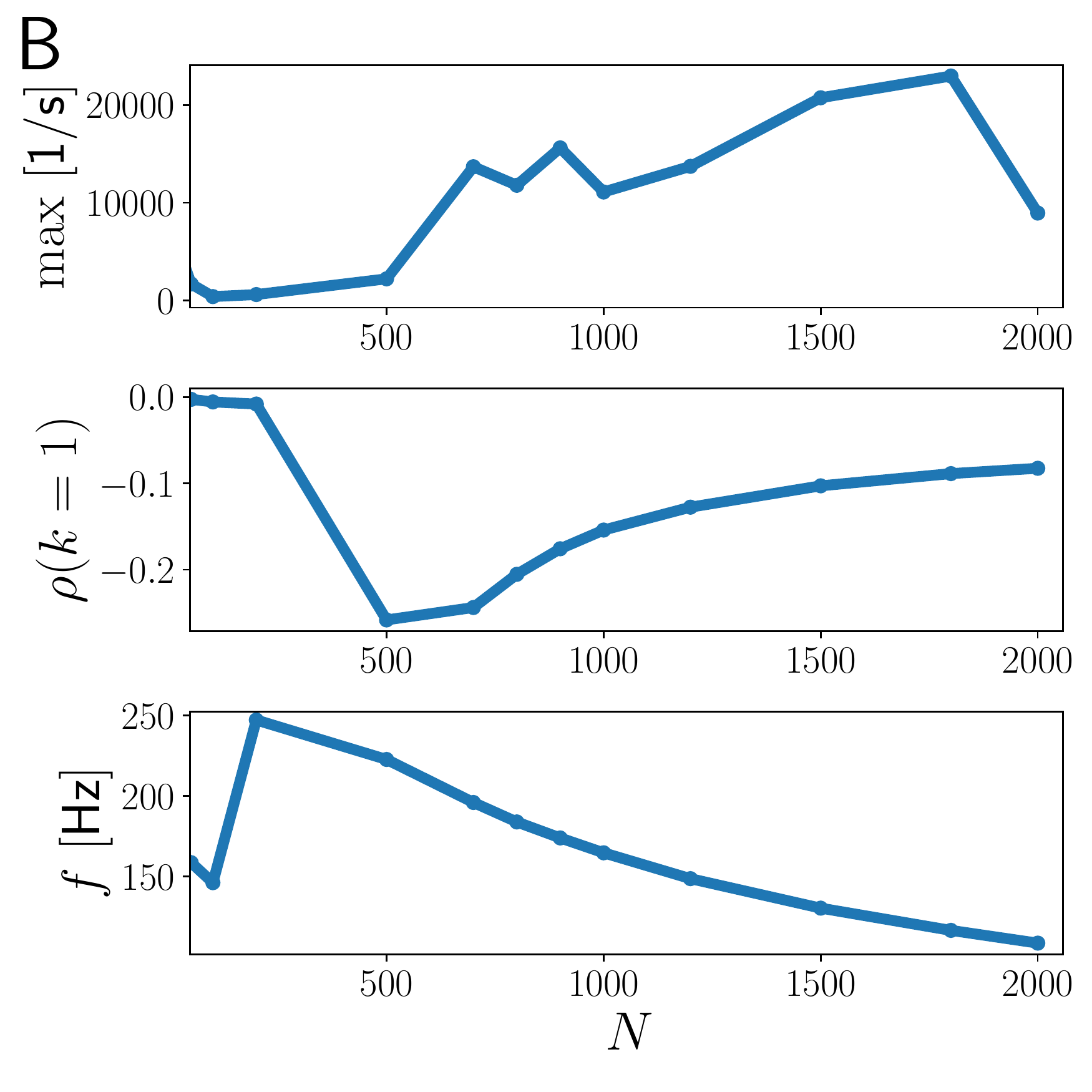}
\caption{Maximum of PSD of network activity, mean network SCC and network frequency $f$ as a function of $I_{0}$ for $N = 500$ (panel A) and as a function of $N$ for $I_{0} = 50~\text{mV}$ (panel B). This corresponds to the vertical white line and to the horizontal black line in \fig \ref{fig_2}. Thus in both cases (A) and (B), the transition correlates with the onset of increase in power of the network oscillation. $d = 200~\text{s}$ simulation time.}
\label{fig:coherence_frequency}
\end{figure}

In \fig \ref{fig:coherence_frequency} A, we plot the value of the PSD at its maximum (which is attained at the frequency $f$ of the network oscillation) as a function of $I_{0}$, together with the frequency $f$ of the network oscillation and the mean network SCC. This plot corresponds to the white vertical line in \fig \ref{fig_2} A. The maximal value of the PSD serves as a measure of coherence for the network oscillation. We see that as the mean network SCC decreases towards negative values, the coherence of the network oscillation increases. The frequency $f$ of the global network oscillation increases only slightly with $I_{0}$. In \fig \ref{fig:coherence_frequency} B (which corresponds to the black horizontal line in \fig \ref{fig_2} A), we show analogous results for the PSD maximum as a function of $N$ for fixed $I_{0}$. 
For larger network sizes, the network frequency $f$ decreases monotonically with $N$, whereas the peak value of the PSD increases. Again, a minimum of the mean network SCC (observed at $N = 500$) is accompanied by an intermediate value of the maximum of the PSD of the network activity. 

\begin{figure}[h!]
\includegraphics[width = 0.4\textwidth]{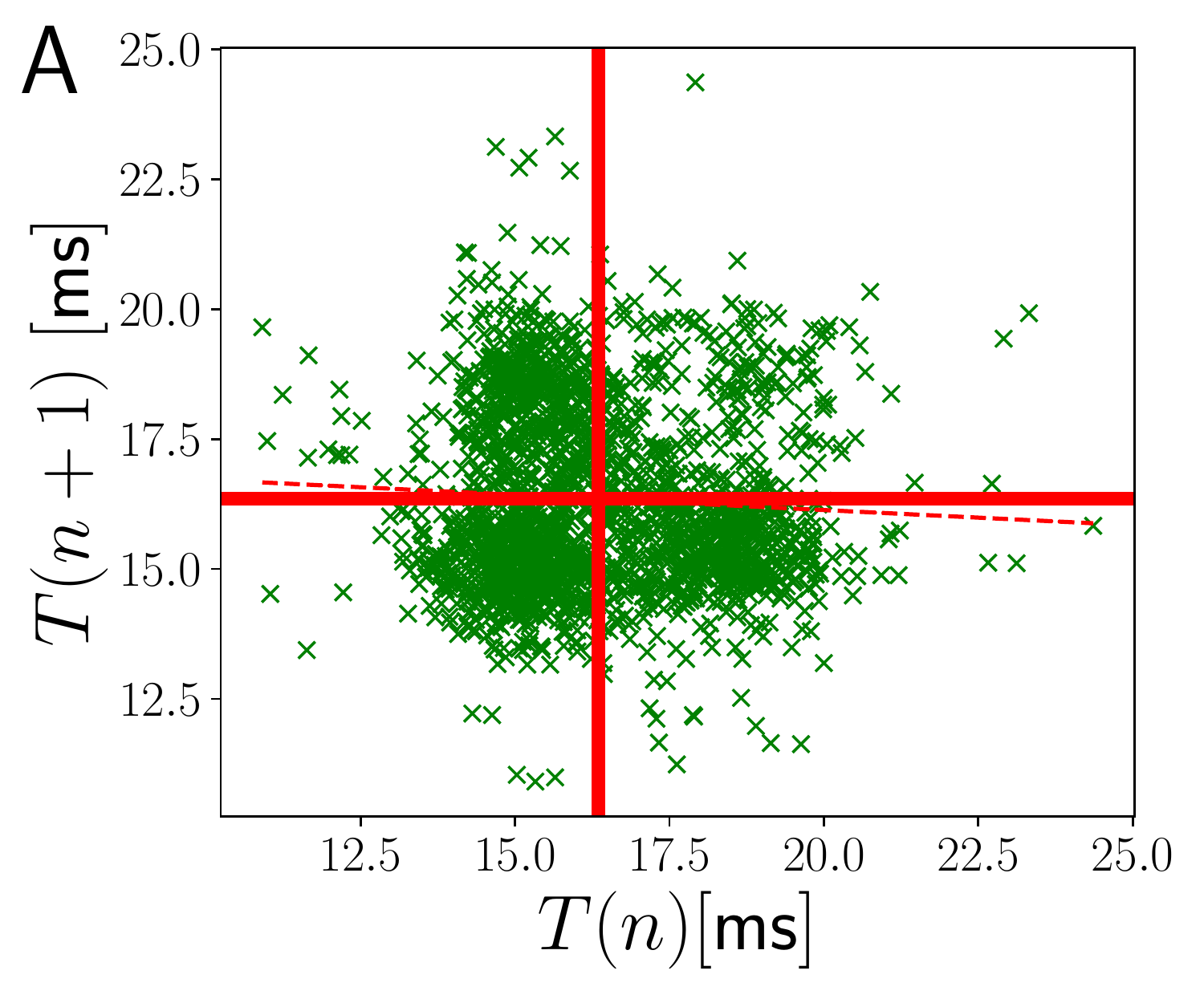}
\includegraphics[width = 0.4\textwidth]{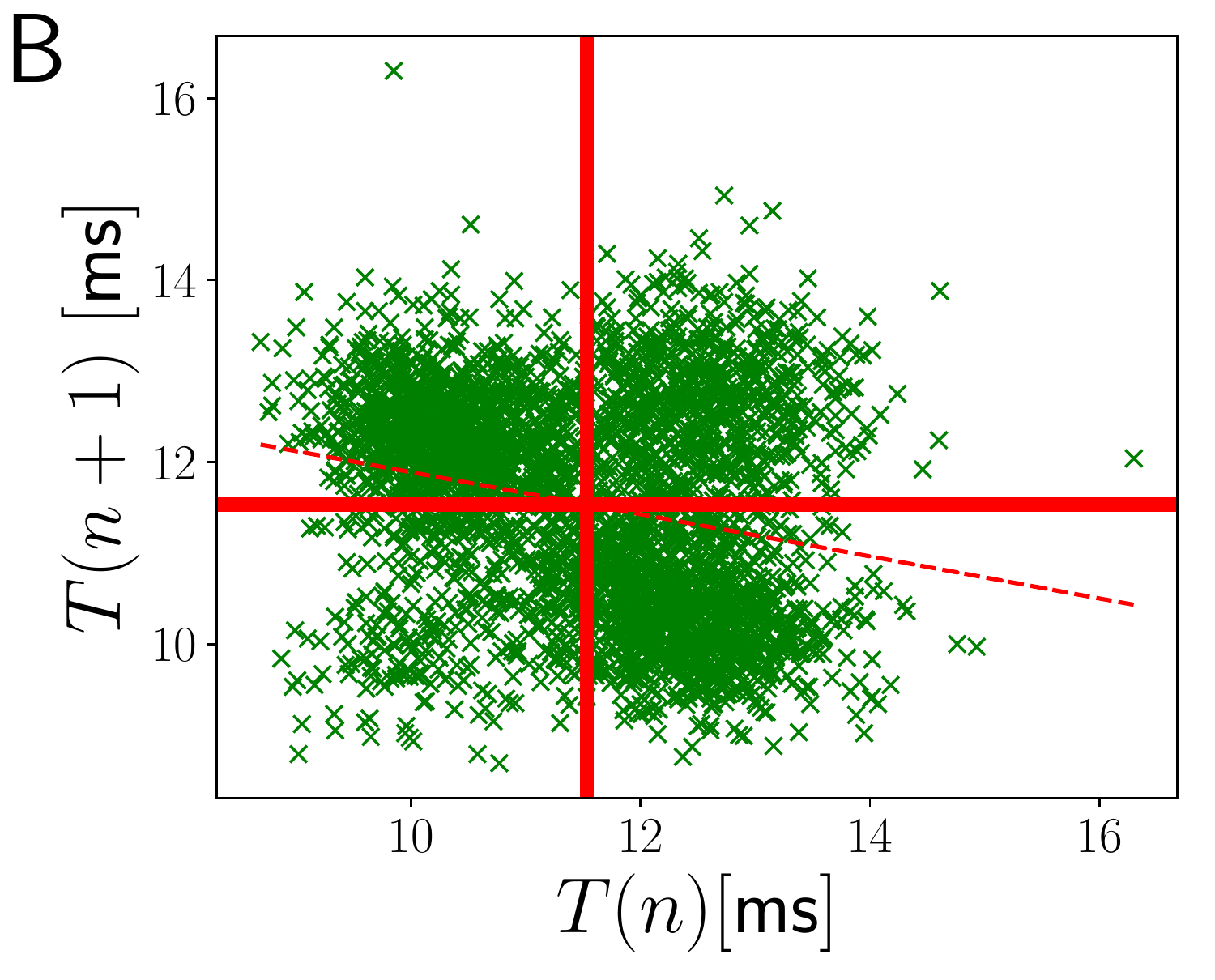}
\includegraphics[width = 0.4\textwidth]{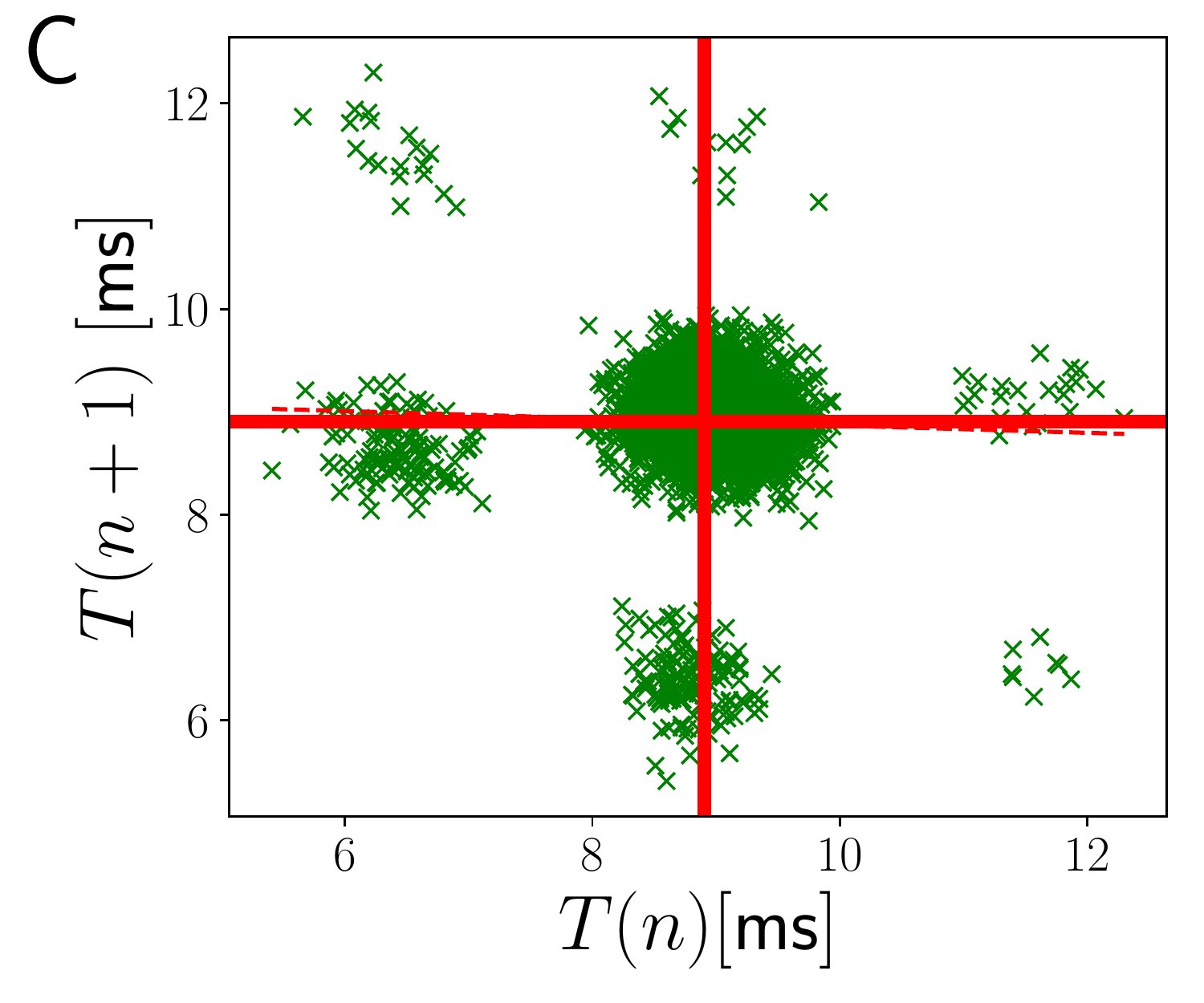}
\caption{Plot of successive ISIs ($n+1$st ISI plotted as function of $n$th ISI) for three different values of $I_{0}$, corresponding to the three white stars in \fig \ref{fig_2}. A: $I_{0} =40~\text{mV}$, A: $I_{0} =50~\text{mV}$, C: $I_{0} =60~\text{mV}$. The red solid vertical and horizontal lines denote the mean ISI. The dashed red line is a linear regression, whose slope is the SCC at lag $1$, which is given by -0.06, -0.23, -0.03 for panels A, B, C, respectively. These values are close to the mean network SCC. Other parameter values, except duration of simulation ($d = 50~\text{s}$), are as in \fig \ref{fig_2}.}
\label{fig:ISI_sketch}
\end{figure}

To determine the microscopic properties of the spike trains at the transition to negative SCCs, we plot consecutive ISIs in \fig \ref{fig:ISI_sketch} for three different values of $I_{0}$, corresponding to the white stars in \fig \ref{fig_2} A. The slope of a linear regression function in these plots then is the SCC at lag $1$ \cite{avila_akerberg_chacron_2011}, which is close to zero for \fig \ref{fig:ISI_sketch} A and C, and negative in panel B. We see in \fig \ref{fig:ISI_sketch} A that all ISIs scatter around the mean value and the SCC stays close to zero. In \fig \ref{fig:ISI_sketch} B, short ISIs tend to be followed by long ones, and vice versa, so that the SCC becomes negative. We checked that the slope is not negative due to the three large outlier ISIs. In \fig \ref{fig:ISI_sketch} C, most ISIs cluster again around the mean value, but there are also smaller clusters reflecting the strongly driven state of the system in which more complex firing patterns, reflecting higher-order mode locking, can occur.

As such, the mechanism for the onset of negative mean network SCCs is similar to the onset of negative ISI correlations in single perfect IF neuron models with another form of negative activity-dependent feedback, namely a spike-triggered adaptation current with a single exponential time scale \cite{schwalger_lindner_PRE_2015, schwalger_13}. In these models, the single-neuron SCC becomes maximally negative at an intermediate value of the ratio between the timescale of the adaptation current providing negative feedback and the deterministic mean ISI of the neuron. In our network setup, the bias drive $I_{0}$ sets the time scale for single-neuron firing, whereas the global rhythm is mainly determined by network properties, in particular the synaptic transmission delay $D$. The analogy, however, is not perfect as firing of the single neuron determines the network rhythm, which in turn determines the negative feedback.

\begin{figure}[h!]
\includegraphics[width = 0.5\textwidth]{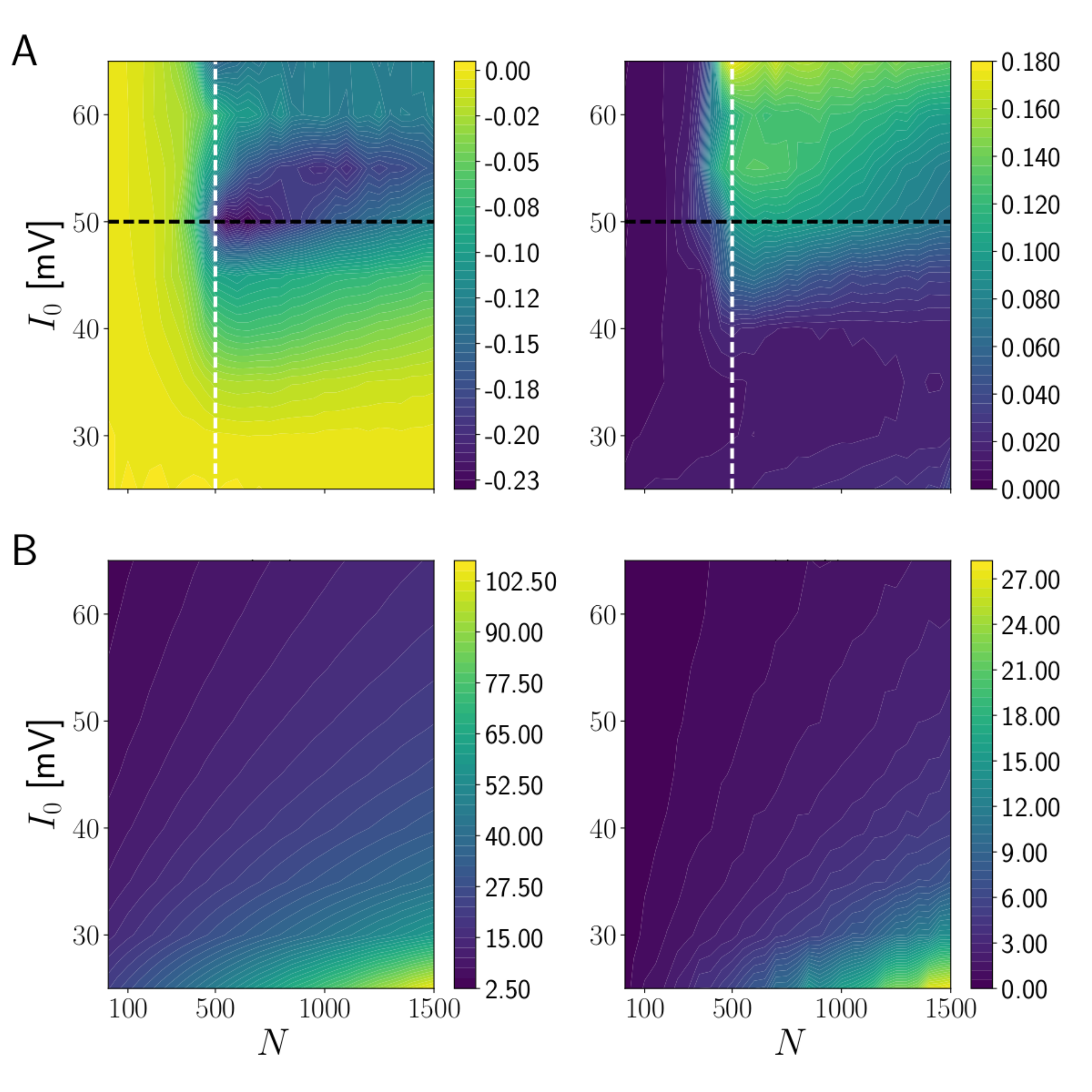}
\caption{A fluctuating number of synapses across neurons in networks with current-based synapses decreases the mean network SCC below zero for a larger range of parameters.
The transition (panel A left) to negative mean network SCCs in this case is accompanied by an increase in the standard deviation of the SCC (panels A right) across neurons.
Mean network SCC (\eq \ref{eq:network_scc}) (\textbf{A}, left) and ISI (\eq \ref{eq:network_ISI}) (\textbf{B}, left) together with standard deviations of SCC (A, right) and of the mean ISI distribution across neurons (B, right) for the $P$-fixed connectivity scenario. The average number of synapses onto a neuron is given by $\overline{C} = p(N-1)$ with $p=0.2$. No refractory period.}
\label{fig_4}
\end{figure}

\begin{figure}[!h]
\includegraphics[width = 0.5\textwidth]{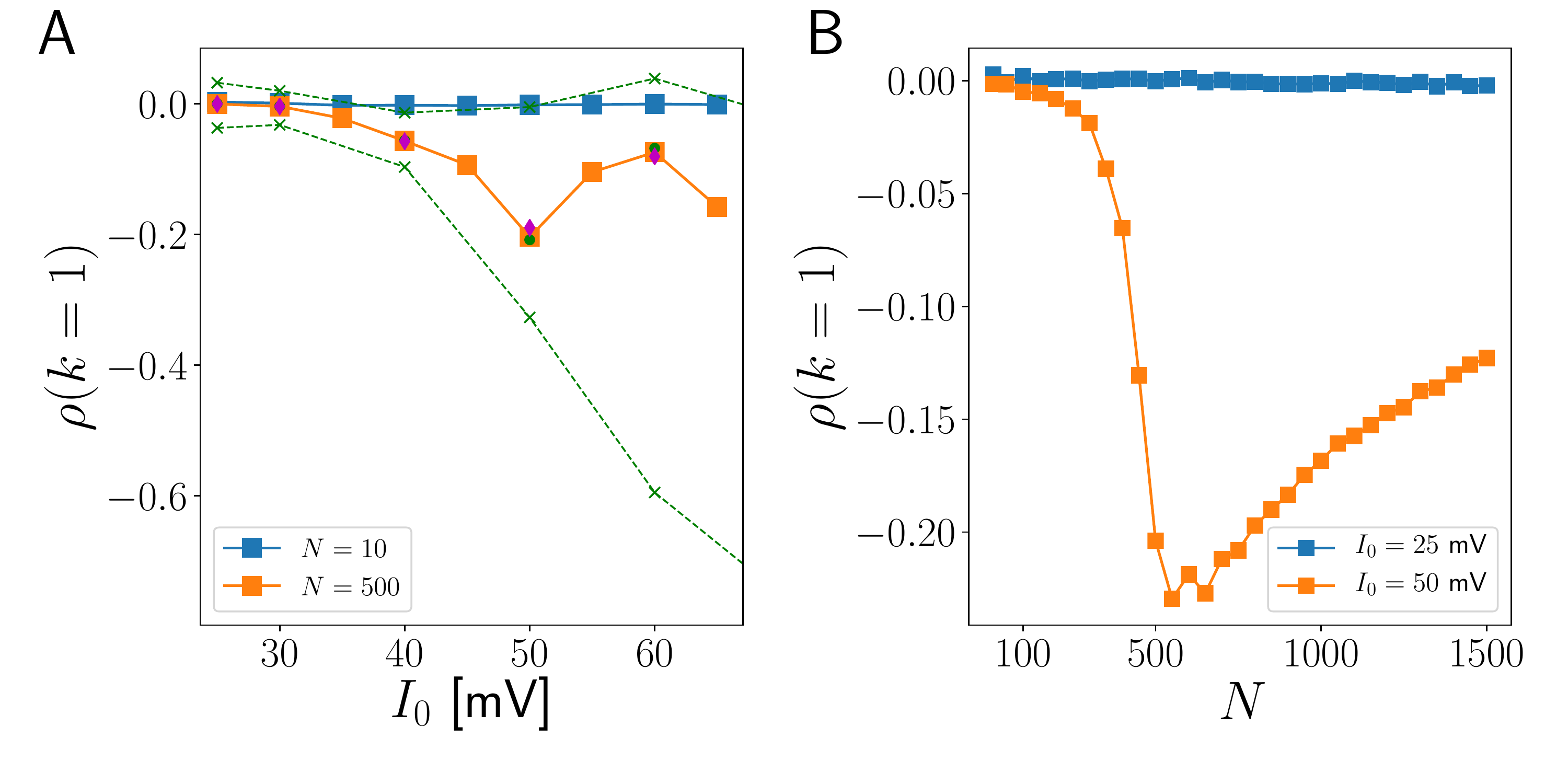}
\includegraphics[width = 0.5\textwidth]{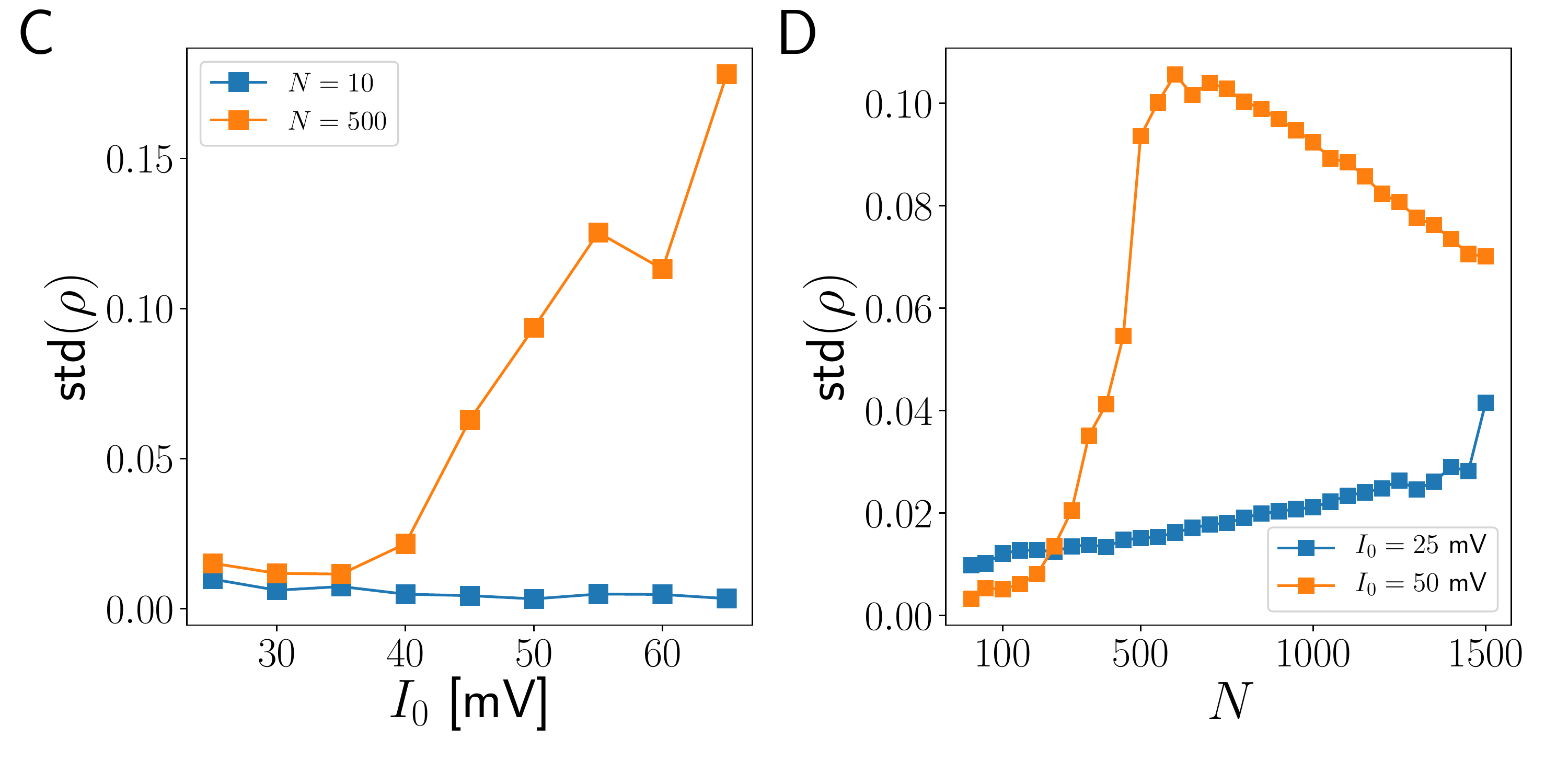}
\caption{Correlation-decorrelation transitions for current-based networks in the $P$-fixed scenario. The mean network SCC (panels A, B) deviates from zero as the standard deviation of the SCC across neurons (panels C, D) increases.
Lineout of \fig \ref{fig_4} A along the white dashed vertical line (\textbf{A}, orange line) and along the black dashed horizontal line (\textbf{B}, orange line). The green symbols are a control simulation started with a different random seed compared to the simulation shown in orange, and with different network realizations for each value of $I_{0}$. The green dashed lines are the minimum and the maximum of the SCC distributions for this simulation. The magenta diamonds are another control simulation with exactly the same network connectivity for each value of $I_{0}$. Hence, the rugged dependence of the mean network SCC for $N = 500$ is not an artifact of finite simulation time or the setup of the network connectivity. C,D: Ensemble standard deviations of the SCC distribution (lineouts of \fig \ref{fig_4} A right panel along the white and black dashed lines). The blue squares are for a smaller value of $N$ (panels A and C) or $I_{0}$ (panels C and D) before the transition to negative mean network SCCs.}
\label{fig_5}
\end{figure}

We now show that the transition to negative mean network SCCs also occurs in networks with a more biologically realistic fluctuating number of synapses ($P$-fixed case, \fig \ref{fig_4} A). Here, in contrast with the $C$-fixed scenario, the standard deviation of the SCC computed across neurons increases at the onset of negative mean network SCCs (\fig \ref{fig_4} B), which is also shown in the lineouts in \fig \ref{fig_5} C and D. Overall, the temporal correlations are slightly less negative than in the $C$-fixed scenario shown in \fig \ref{fig_2}. The mean network ISI is generally larger than in the $C$-fixed case, i.e. the average neuron is subject to more inhibition than in the $C$-fixed case, which at the same time slightly reduces the mean network SCC for fixed values of $N$ and $I_{0}$.

\subsubsection{Effect of refractory period}
\begin{figure}[h!]
\includegraphics[width = 0.5\textwidth]{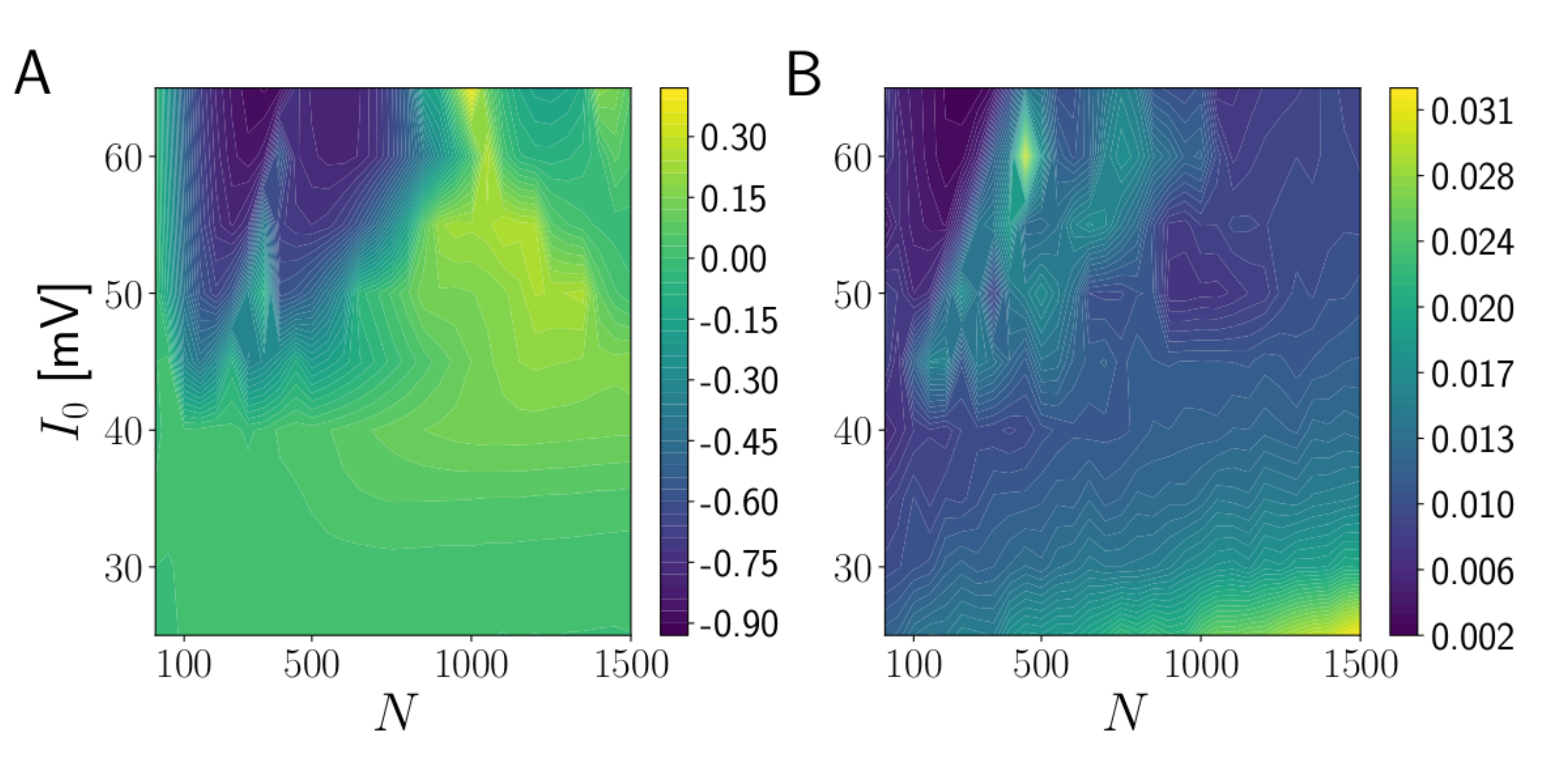} 
\caption{Mean network SCC (A) and its standard deviation across neurons (B) in the presence of a refractory period in the $C$-fixed scenario. The parameter values are like in \fig \ref{fig_2}, but with a refractory period $\tau_{r} = 2~\text{ms}$.}
\label{fig_6}
\end{figure}

In \fig \ref{fig_6}, we show that for the $C$-fixed scenario in the presence of an absolute refractory period, the mean network SCC transitions to small negative values for smaller $N$ as $I_{0}$ is increased, and to intermediate positive values for larger $N$. The critical value for $I_{0}$ for the generation of negative mean network SCCs remains similar to the case without a refractory period, however, the transition is observed at smaller values of $N$.

\subsection{SCC in networks with conductance-based synapses}

\begin{figure}[h!]
\includegraphics[width = 0.5\textwidth]{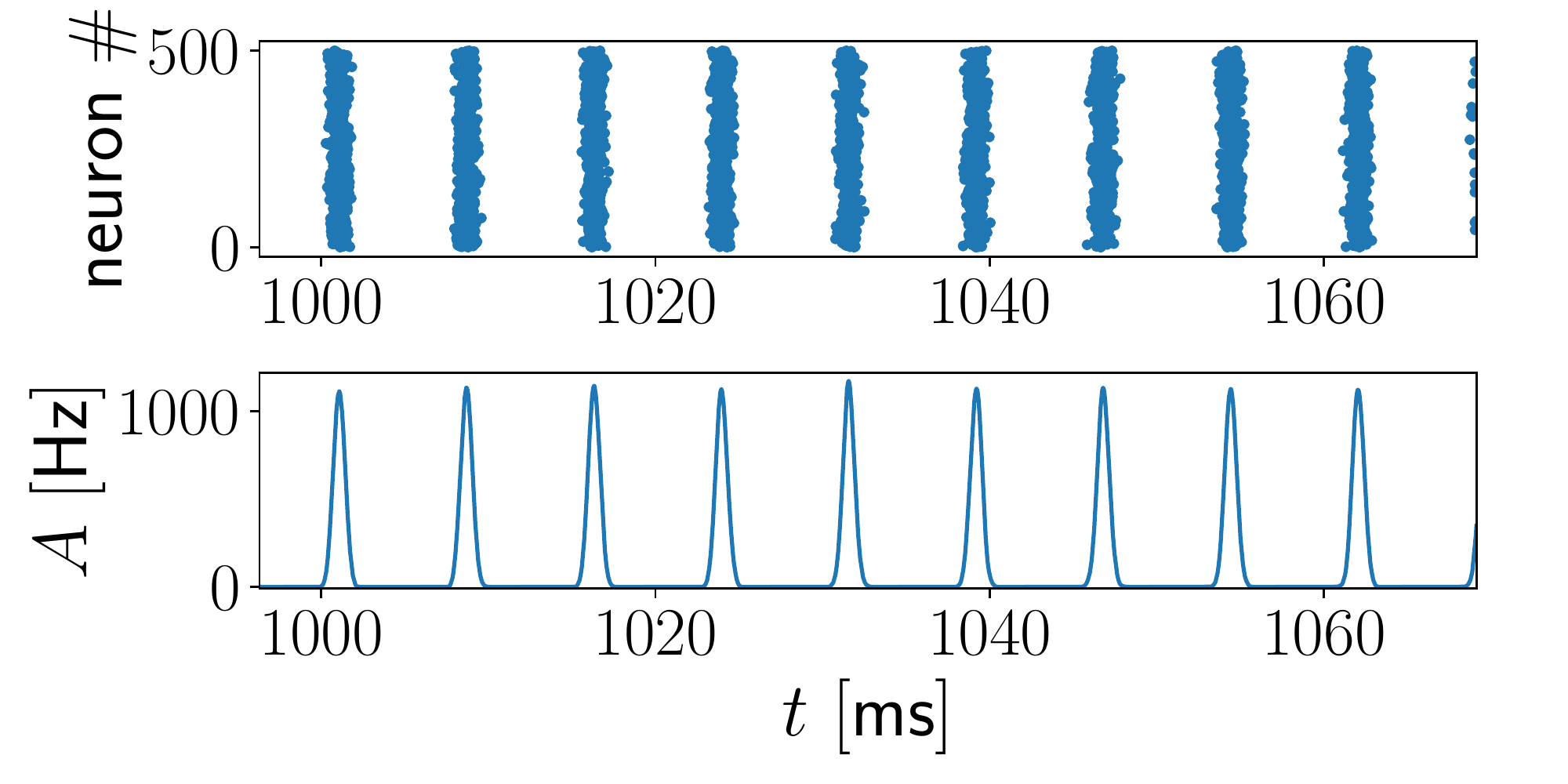}
\includegraphics[width = 0.5\textwidth]{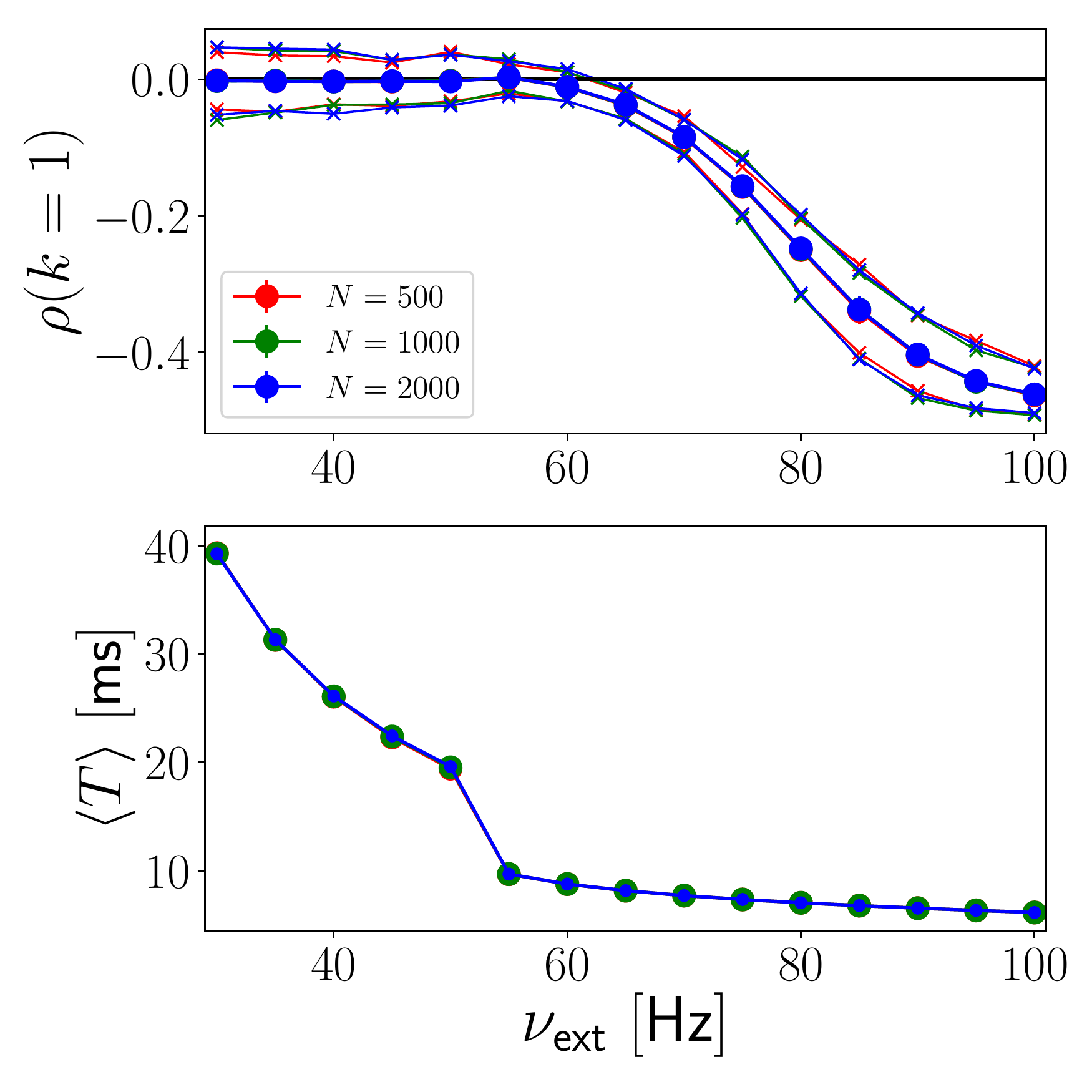} 
\caption{Transition to negative mean network SCCs in a network of conductance-based IF neurons with increasing external drive. \textbf{Top:} Network activity at $\nu_{\text{ext}} = 70~\text{Hz}$. The network oscillation has a high frequency $f \approx 140~\text{Hz}$. \textbf{Bottom:} Mean network SCC and ISI. Thin lines with crosses show minimum and maximum of the SCC distribution.  $C$-fixed scenario with $pN = 25$, corresponding e.g. to $p = 0.05$ for $N = 500$. Parameter values: $n^{\text{P}}_{I}=800$, for remaining parameters, see Appendix \ref{sec:coba}. }
\label{fig_7}
\end{figure}

The mechanism for the generation of negative temporal correlations relies on the onset of a coherent network oscillation in a simplified model of neural dynamics. In particular, the current-based synapses are instantaneous and do not follow biologically realistic dynamics. We therefore investigated whether transitions to negative mean network SCCs could also be observed in more biologically plausible neuronal networks capable of fast oscillations.
To that end, we considered networks of inhibitory LIF neurons with single-neuron parameters similar to those for fast-spiking basket cells in region CA1 of hippocampus \cite{donoso_et_al_2018}. The model is described in detail in Appendix \ref{sec:coba}. The refractory period is always set to $1~\text{ms}$ in the following plots. Instead of a constant bias current $I_{0}$, we drive every neuron with excitatory Poisson processes, whose strength (measured by their constant rate $\nu_{\text{ext}}$) determines the coherence of a nascent network oscillation. The frequency $f$ of the network oscillation is largely independent of the Poisson drive and depends mainly on the synaptic parameters \cite{brunel_wang_2003}. 
For a network oscillation in the fast gamma range, the mean network SCC transitions to negative values at a value of $\nu_{\text{ext}} \approx 60~\text{Hz}$ (\fig \ref{fig_7}), well in the suprathreshold regimes, which requires $\nu_{\text{ext}} \approx 13~\text{Hz}$ ,as determined by numerical simulation.

\begin{figure}[h!]
\includegraphics[width = 0.5\textwidth]{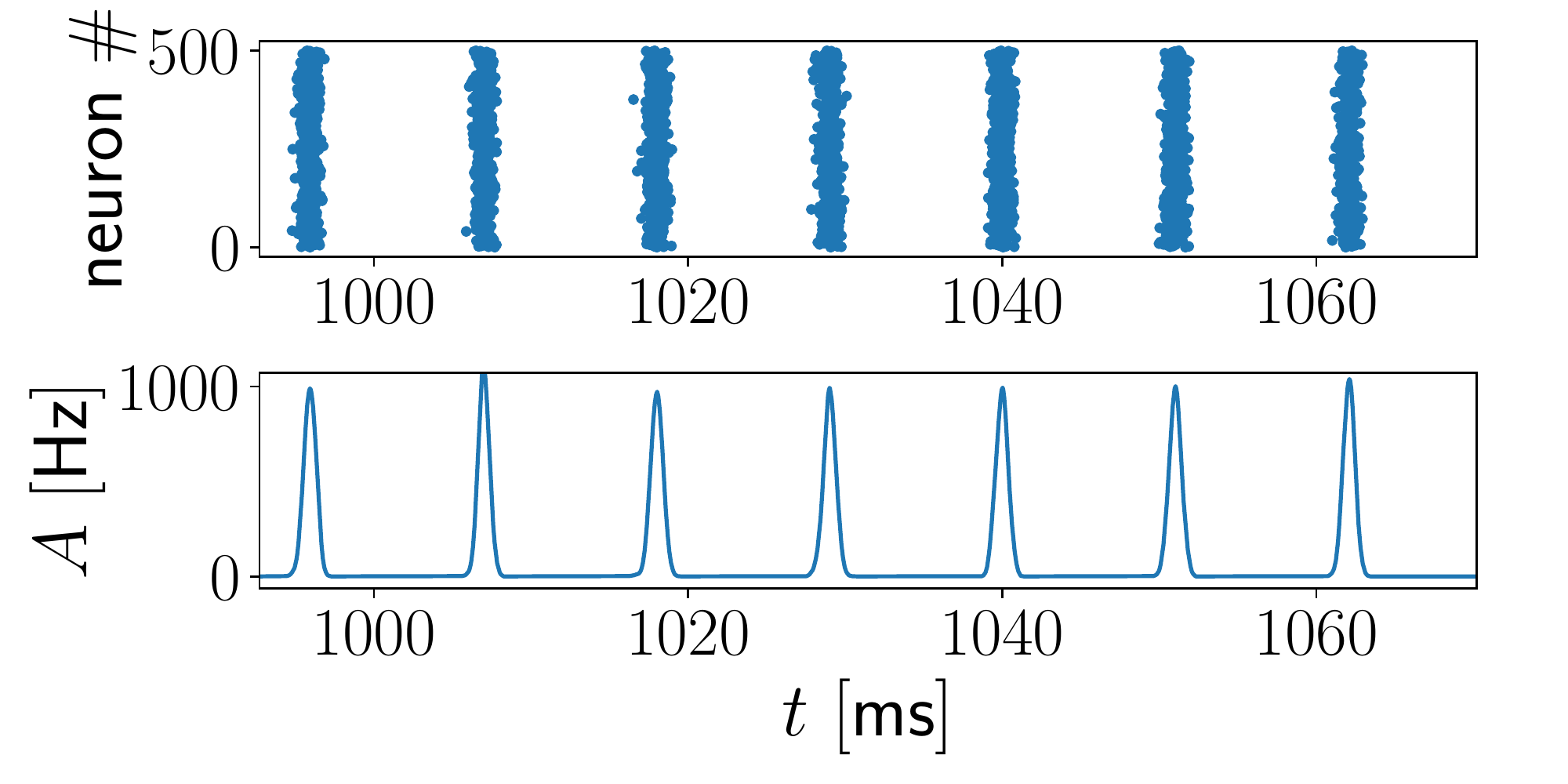}
\includegraphics[width = 0.5\textwidth]{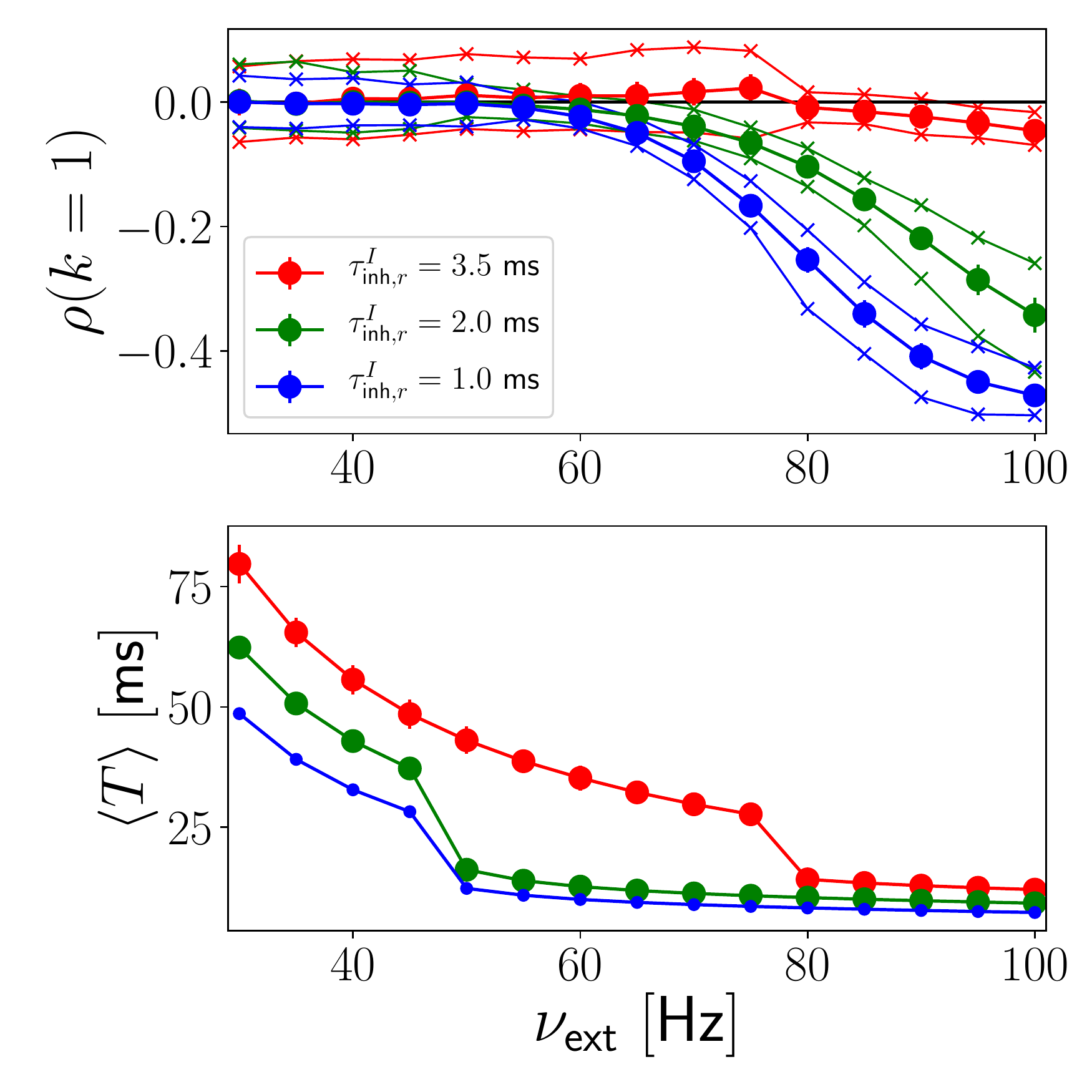} 
\caption{Transition to negative mean network SCCs in network of conductance-based IF neurons with increasing external drive. \textbf{Top:} Network activity at $\nu_{\text{ext}} = 70~\text{Hz}$ for $\tau^{I}_{\text{inh},r} = 2.0~\text{ms}$. The network oscillation has a lower frequency $f \approx 80~\text{Hz}$. \textbf{Bottom:} Mean network SCC and ISI. Thin lines with crosses show minimum and maximum of the SCC distribution.  $N = 500$.  $C$-fixed scenario with $p = 0.05$. Parameter values: $n^{\text{P}}_{I}=800$, for remaining parameters, see Appendix \ref{sec:coba}. }
\label{fig_8}
\end{figure}

Can transitions to negative mean network SCCs still be observed in the presence of a slower oscillation in the gamma range? We increased the synaptic rise time from $\tau^{I}_{\text{inh}, r} = 0.45~\text{ms}$ in \fig \ref{fig_7} to three higher values in \fig \ref{fig_8}, which reduces the network frequency \cite{brunel_wang_2003}. We observe transitions to negative mean network SCCs in \fig \ref{fig_8}, which are strongest for  for $\tau^{I}_{\text{inh}, r}$ on the order of the refractory period. For $\tau^{I}_{\text{inh}, r} \geq 3.5~\text{ms}$, the network activity does not oscillate coherently anymore and therefore, the mean network SCC does not decrease as much as for smaller $
\tau^{I}_{\text{inh}, r}$.

\begin{figure}[h!]
\includegraphics[width = 0.5\textwidth]{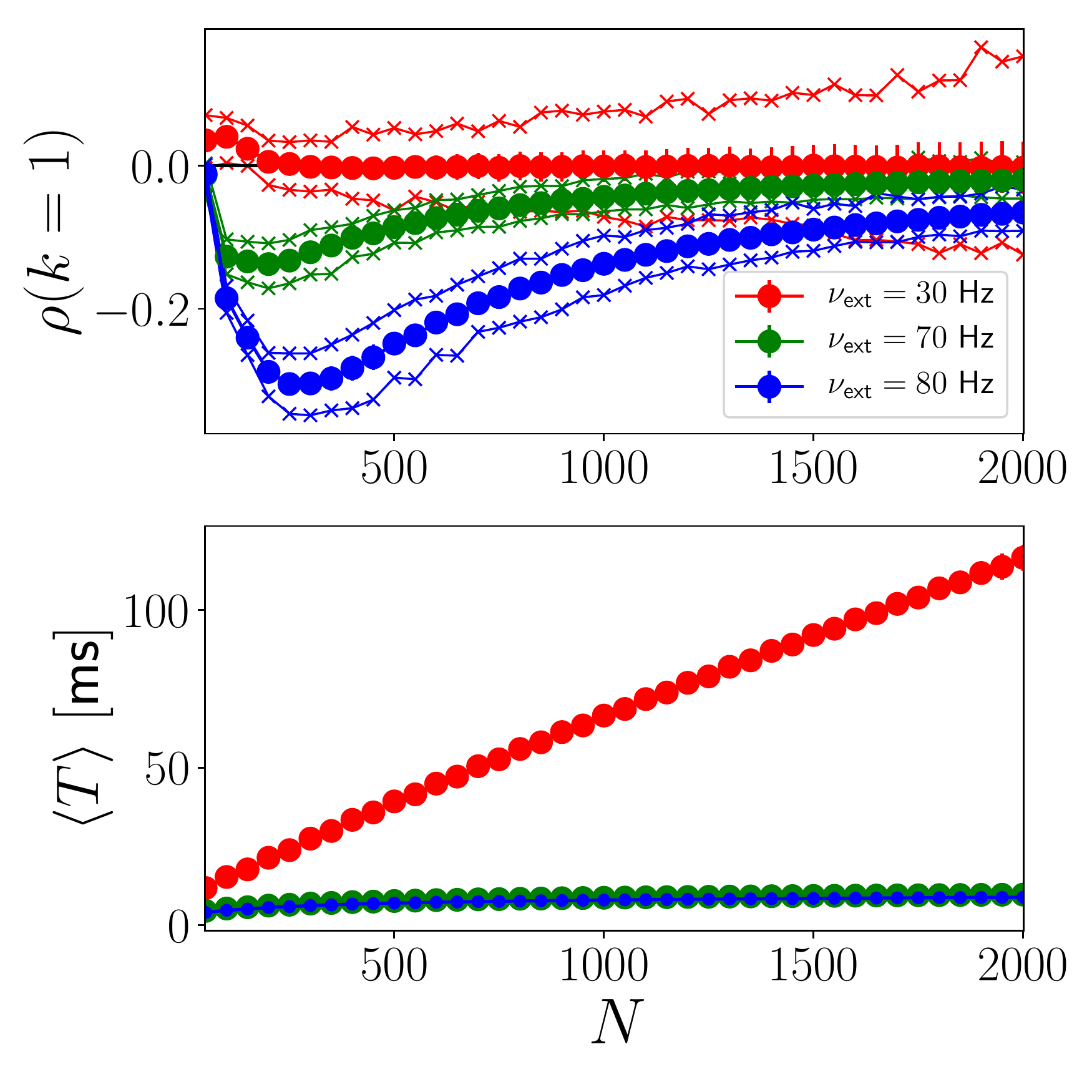} 
\caption{Transition to negative mean network SCCs in network of conductance-based IF neurons with increasing network size $N$. $C$-fixed scenario with $p = 0.05$. Thin lines with crosses show minimum and maximum of the SCC distribution.  For $N=500$, the network oscillation has the same frequency as in \fig \ref{fig_7} at the corresponding values of $\nu_{\text{ext}}$. Parameter values: $n^{\text{P}}_{I}=800$, for remaining parameters, see Appendix \ref{sec:coba}. }
\label{fig_9}
\end{figure}

Finally, we show that as in current-based networks, the mean network SCC also transitions to negative values as the network size is increased (\fig \ref{fig_9}), although there is seemingly no flat part near a mean network SCC close to zero before the transition to negative values unless it happens at low $N$ , as in e.g. \fig \ref{fig_4} B- the transition to negative mean network SCCs as a function of $N$ occurs faster than in current-based networks.

Taken together, we have shown that at the onset of a coherent global network oscillation, neural networks with both current- and conductance-based synapses generate negative ISI correlations in the output spike trains of their constituent neurons, with a parameter dependence that is generally not monotonic.

\section{Noise-reduced diffusion approximation}

\label{sec:DA}
Which features of the statistics of the input spike train (cf. \eq \ref{eq:synaptic_current}) to a given neuron causes its output spike to show non-renewal statistics?

One way to address this question is to consider the seminal diffusion approximation (DA). In the DA for purely inhibitory networks, the synaptic current fed into one average ''typical`` neuron in the network is given by \eq \ref{eq:diffusion_approximation}. Assuming that the coupling between neurons is sparse, pairwise correlations between neurons can be neglected. This entails that the statistics of one neuron are representative of the whole neuronal network. With some additional technical assumptions, a mean-field type description of the dynamics is possible \cite{brunel_hakim_1999}, because it is possible to write a one-dimensional time-dependent Fokker--Planck equation (FPE) for the evolution of the membrane potential of one single cell. The formulation of the FPE together with its boundary conditions is self-consistent, because the solution of the FPE gives the time-dependent firing rate $\nu(t)$ of the neuron; this quantity also appears in the FPE via the time-dependent drift and diffusion coefficients \eqs \ref{eq:drift_term} and \ref{eq:diffusion_term}, respectively.
If the network is sparse, this approach means that the output spike train of any given neuron in the network and its input spike train share the same statistics. This was used in \cite{dummer_wieland_lindner_2014} to self-consistently determine spike train power spectra in networks of EI neurons using an iterative numerical algorithm in the $C$-fixed scenario. This self-consistent approach does not apply, however, if the connectivity is not sparse or the network activity is synchronous.

\subsection{Sensitivity of SCC to the number of input spikes}
\begin{figure}[h!]
\includegraphics[width = 0.5\textwidth]{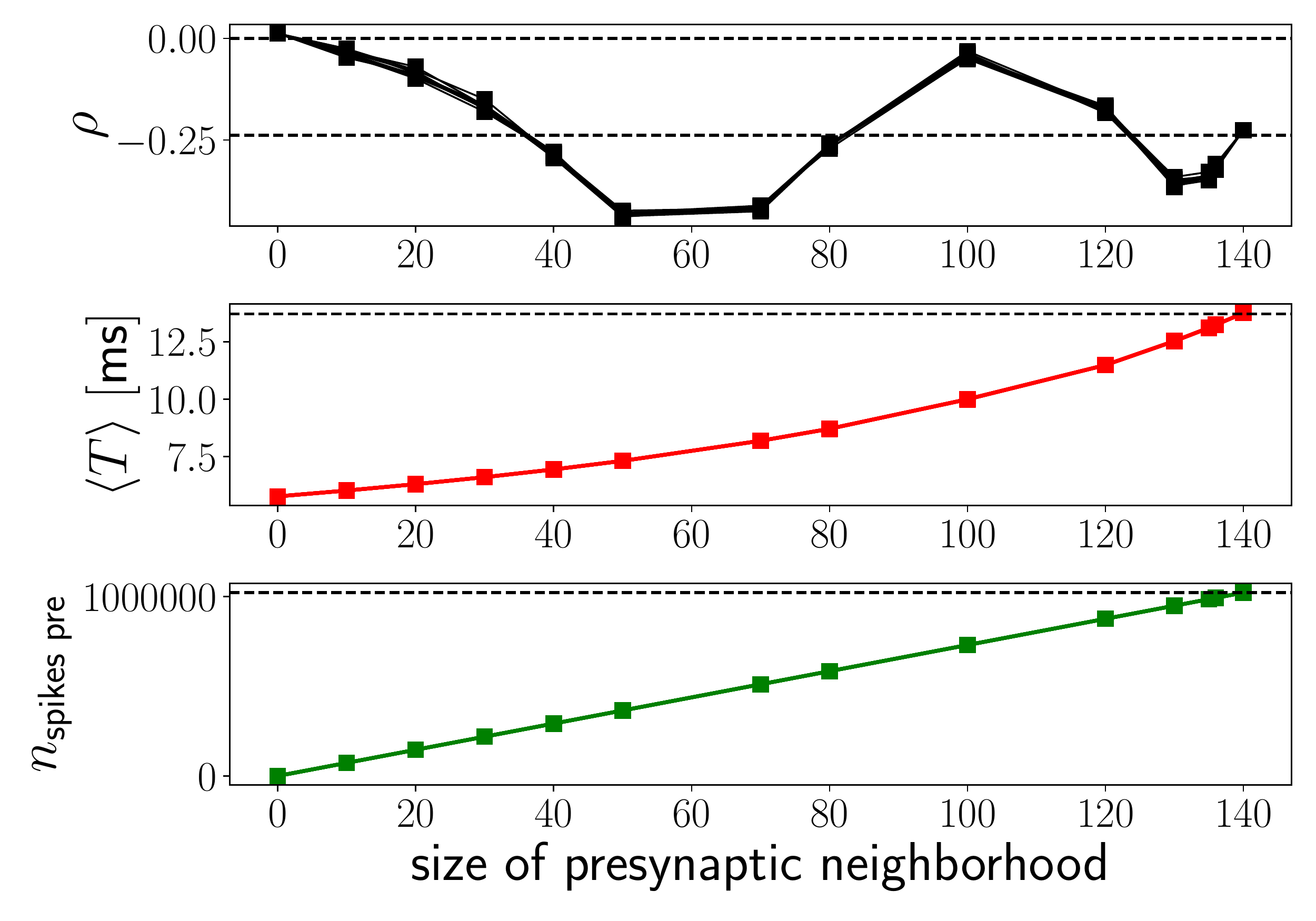}
\caption{The SCC (top panel) of one neuron embedded in the network is very sensitive to the number of presynaptic spikes (bottom panel), in contrast to the monotonically increasing mean ISI (middle panel). The maximal value of $C$ is $140$, and the abscissa is swept from left to right by letting the pres-synaptic neighborhood size increase from $0$ to $140$. The dashed horizontal lines are at an SCC value of zero (upper line) and the online SCC value for the chosen neuron (lower line). The dashed horizontal line in the plot for the mean ISI is the online value of the mean ISI for the chosen neuron. The different lines (hardly distinguishable, because they largely overlap) are for $10$ random permutations of the order of the presynaptic spike trains to ensure that the non-monotonic behavior of the SCC is not a result of a particular arrangement of the order of pre-synaptic spike trains. Parameter values: $N = 700$, $p =0.2$ (C-fixed scenario). $I_{0} = 50~\text{mV}$, $d= 100~\text{s}$ simulation time. Other parameters as in \fig \ref{fig_2}.}
\label{fig_10}
\end{figure}

First, we determined how sensitive the SCC of a neuron embedded in a network is to changes in the size of its pre-synaptic neighborhood (\fig \ref{fig_10}). We recorded all spike trains present in the corresponding pre-synaptic neighborhood and then ran an offline simulation in which the recorded inhibitory spikes were manually fed into the simulation. As the size of the pre-synaptic neighborhood increases, the mean ISI (middle panel of \fig \ref{fig_10}) increases monotonically, in contrast to the SCC, which surprisingly shows non-monotonic behavior with increasing size of the presynaptic neighborhood until it reaches its online value when the size of the presynaptic neighborhood reaches its online value given by $C = p(N-1)$.

Thus, we have verified that feeding back all pre-synaptic spike trains to a chosen neuron in an 'offline' simulation of that neuron reproduced the output statistics obtained during the full network simulation. The behavior of the SCC with increasing pre-synaptic neighborhood size is non-monotonic, and nearly all input spikes (bottom panel of \fig \ref{fig_10}) are needed to reproduce the online value. We have checked that the behavior of the SCC does not depend strongly on which neuron in the network is chosen, which is a result of the low standard deviation of the SCC distribution across neurons for the $C$-fixed case (cf. \fig \ref{fig_2} A right panel).

\subsection{Statistics of the input spike trains impinging on a given neuron}
\begin{figure}[h!]
\includegraphics[width = 0.35\textwidth]{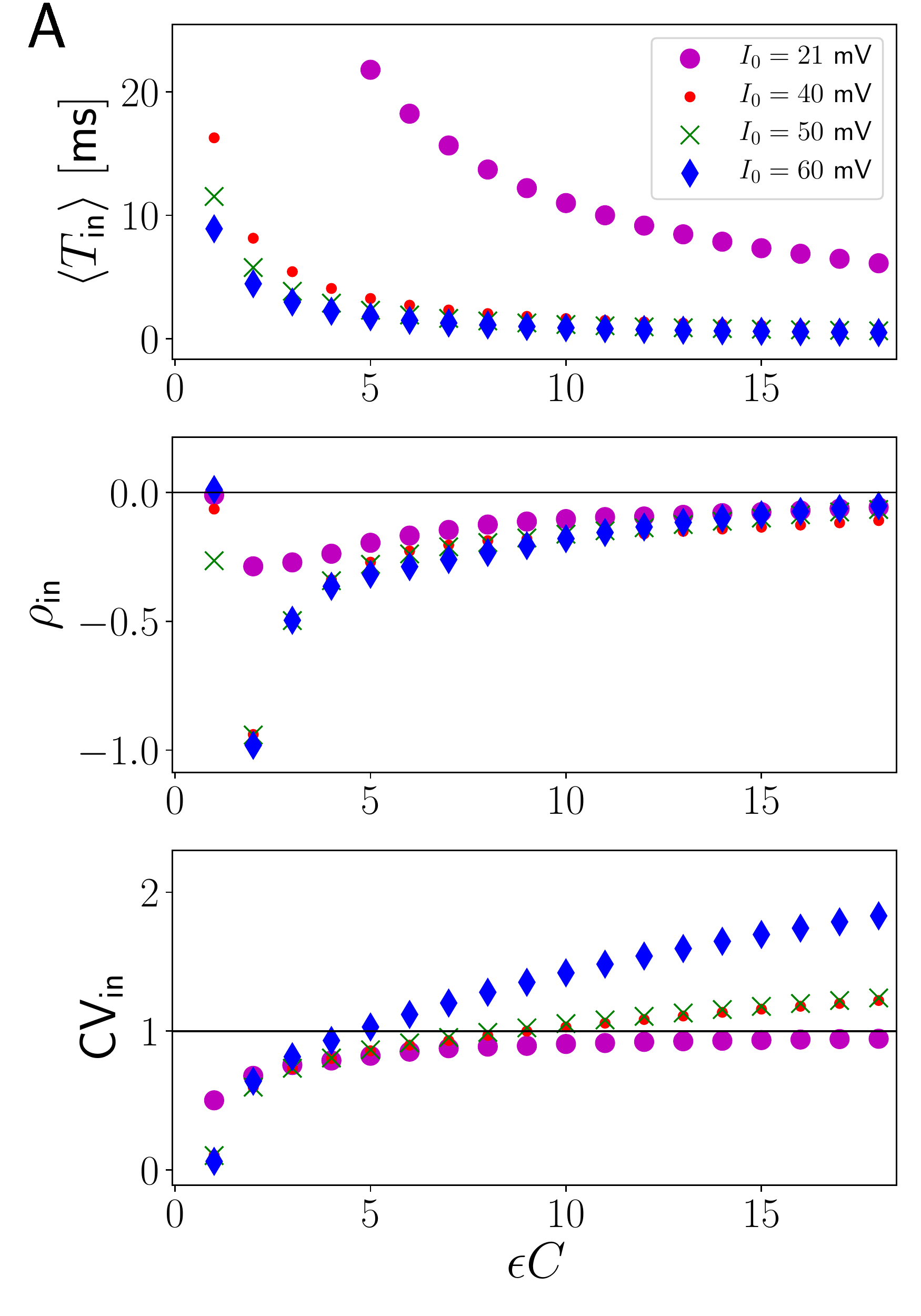}
\includegraphics[width = 0.35\textwidth]{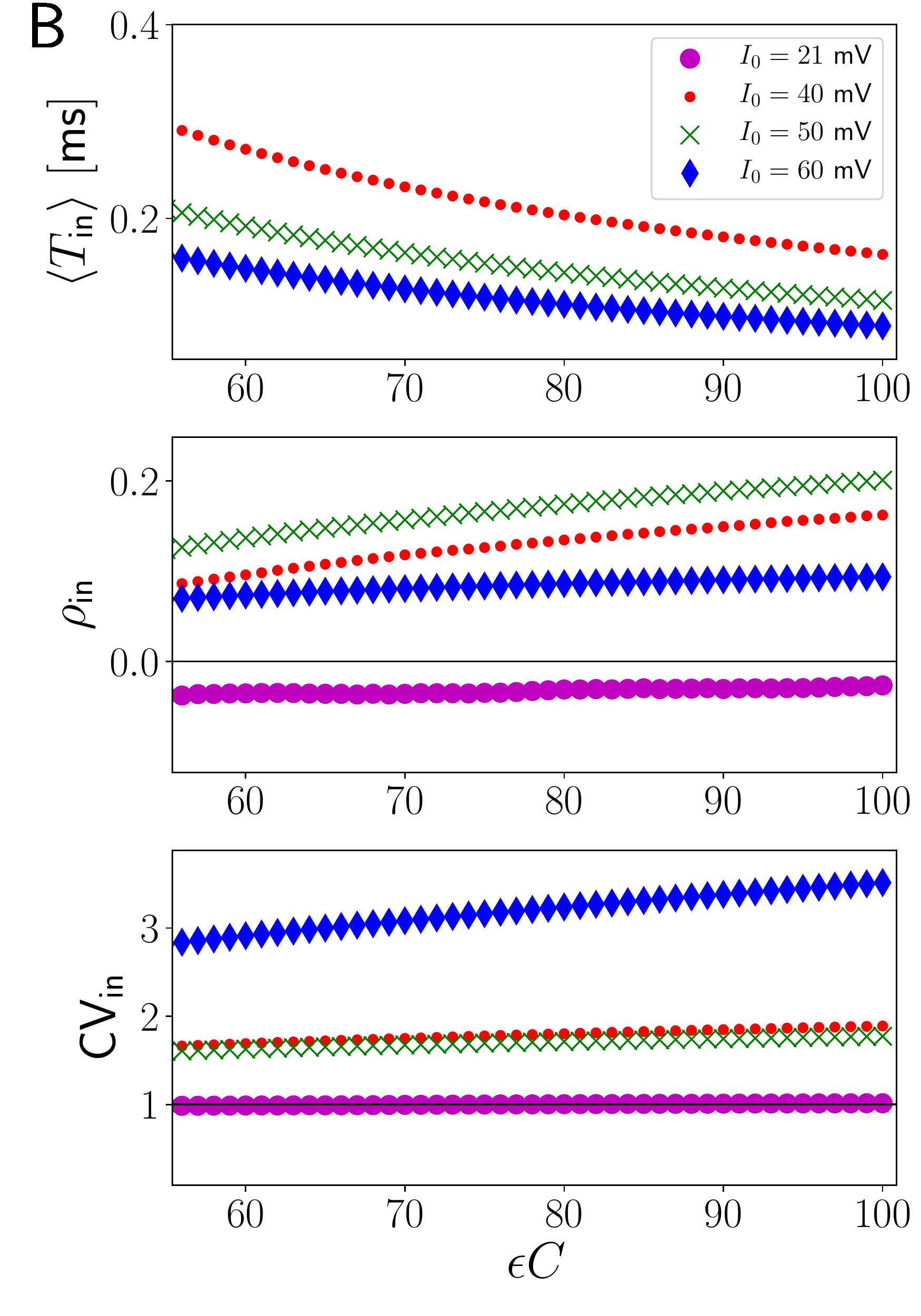} 
\caption{Statistics of pre-synaptic input of one fixed neuron in a network of current-based neurons without refractory period. From top to bottom: input mean ISI, SCC and CV. $\epsilon$ varies between $\frac{1}{C}$ and $1$ from left to right. \textbf{A:} small $\epsilon$. \textbf{B:} large $\epsilon$. Parameters: $N = 500$, other parameter values as in \fig \ref{fig_2}, in particular, $C = p(N-1) = 100$.}
\label{fig_11}
\end{figure}

So far, we have not yet quantified the statistics of the input spike trains. However, for the DA to be applicable, the network has to be in an asynchronous regime, in which neurons discharge according to a Poisson process with time-dependent rate $\nu(t)$. We show the statistics of the pre-synaptic input to one neuron in a network in \fig \ref{fig_11} as a function of the size of its presynaptic neighborhood $\epsilon C$, which is obtained by the superposition of a fraction $\epsilon \in [\frac{1}{C},1]$ of actual pre-synaptic spike trains impinging on the chosen neuron. These pre-synaptic spike trains were recorded in a full network simulation. We also checked that the input statistics did not depend on the chosen neuron in the network, as expected for the $C$-fixed case. In addition to the mean ISI and the SCC, we also computed the coefficient of variation ($\text{CV}$) of the input spike sequence, defined as the ratio between standard deviation and mean of the pooled ISI sequence.

Note that for $\epsilon = \frac{1}{C}$, i.e. when only one input spike train is considered, we recover the results for the output statistics of the neuron.
The mean input ISI (top panel in \fig \ref{fig_11}) decreased with increasing number of pre-synaptic spike trains approximately like $\frac{1}{\epsilon C}$, as expected in the Poisson limit. For small suprathreshold $I_{0}=21~\text{mV}$ (purple circles in \fig \ref{fig_11}), the input CV (bottom panel) is close to $1$ and the input SCC (middle panel) is close to zero, and thus, the network is in a regime where the input spike trains follow Poisson statistics. Increasing $I_{0}$ to larger suprathreshold values (small red circles, green crosses and blue diamonds in \fig \ref{fig_11}, the same values as the three white stars in \fig \ref{fig_2}) results in positive input SCCs and input CVs exceeding $1$, so that the input statistics are then manifestly non-Poissonian. 

\subsection{Noise-reduced diffusion approximation in an offline simulation}
In light of these findings, we now explore whether a DA can still be used to compute both the mean network SCC and ISI. We first describe an open-loop or \textit{offline} simulation scheme to address this question; it uses an adjustable form of the DA for the simulation of one effective neuron in the network. We then compare the results for mean network SCC and ISI obtained during an online full network simulation and the offline effective single-neuron simulation. The algorithm to compare online and offline simulations results is as follows:

\vspace{3pt}
\noindent
1. Simulate full network \textit{online}, and obtain the network activity $A(t)$ (smoothed with rectangular kernel of width $\sigma_{w} = 0.01~\text{ms}$).

\vspace{3pt}
\noindent
2. Simulate one single \textit{offline} neuron (cf. \eq \ref{eq:definition_X}) according to the Langevin equation
 
\begin{align}
\label{eq:DA_alpha_J}
\dot{X} = &\gamma_{t}\left(-X + I_{0} - CJ\nu(t-D) \tau_{t} \right) \\
 &+ \gamma_{t} \sqrt{\alpha_{J} J^{2} C \nu(t-D)\tau_{t} + \sigma_{0}^{2}}\sqrt{\tau_{t}} \xi(t) \nonumber \, ,
\end{align}
 
with a timestep $h = 10^{-2}~\text{ms}$ using the DA with $\nu(t) = A(t)$ in \eq \ref{eq:diffusion_approximation} as drive. We introduced a factor $\alpha_{J}$ that scales the contribution of the network activity $\nu$ to the noise strength in the DA. As for the full network simulations, the Euler-Maruyama method is used for the integration of the stochastic differential equation \ref{eq:DA_alpha_J}.

\vspace{3pt}
\noindent
3. Compute mean SCC at lag $1$ and its standard deviation for the network.

\vspace{3pt}
\noindent
4. Compute SCC at lag $1$ for the chosen offline single neuron.

\vspace{3pt}
\noindent
5. Compare the SCC result from online network simulation and offline single-neuron simulation.

\vspace{5pt}

Another way for estimating the mean network SCC is to simulate an inhomogeneous Poisson process (iPP) \cite{cox_lewis_1966, lewis_shedler_1978} with the same time-varying smoothed rate $A(t)$ as the full network and then compute the mean ISI and the SCC for the generated spike train. We simulated the iPP using the numerical scheme proposed in \cite{laub_et_al_2015}. The idea is that this simple surrogate for the network activity, which has temporal correlations solely due to its time-varying rate (i.e. there are no intrinsic correlations), can give insight into the behavior of the SCC we observe below.

We will now only consider the $C$-fixed scenario, since the DA should then reproduce the behavior of one typical neuron accurately. It would also be possible to compute the mean network SCC in the $P$-fixed scenario by averaging SCC results over the (binomial) distribution of the C-value, a computationally expensive task. We verified that the choice of integration timestep $h$ of the single neuron dynamics as well as the smoothing width $\sigma_{w}$ do not impact the following results (see \fig \ref{fig:fig_14} below). 

In \fig \ref{fig_12}, we show the performance of the DA and iPP approximation schemes for both the mean network SCC (top panels) and ISI (bottom panels) as $I_{0}$ is increased (cf. the vertical white dashed line in \fig \ref{fig_2}). For small $I_{0}$ before the transition to negative mean network SCCs, both approximation schemes agree well with the full network simulations. At the onset of negative mean network SCCs, both the DA and the iPP approximation schemes fail to reproduce the negative mean network SCC of the full online simulation, although the DA shows a slight decrease of the SCC. The mean network ISI, in contrast, is always approximated well by both schemes.

\begin{figure}[h!]
\includegraphics[width = 0.4\textwidth]{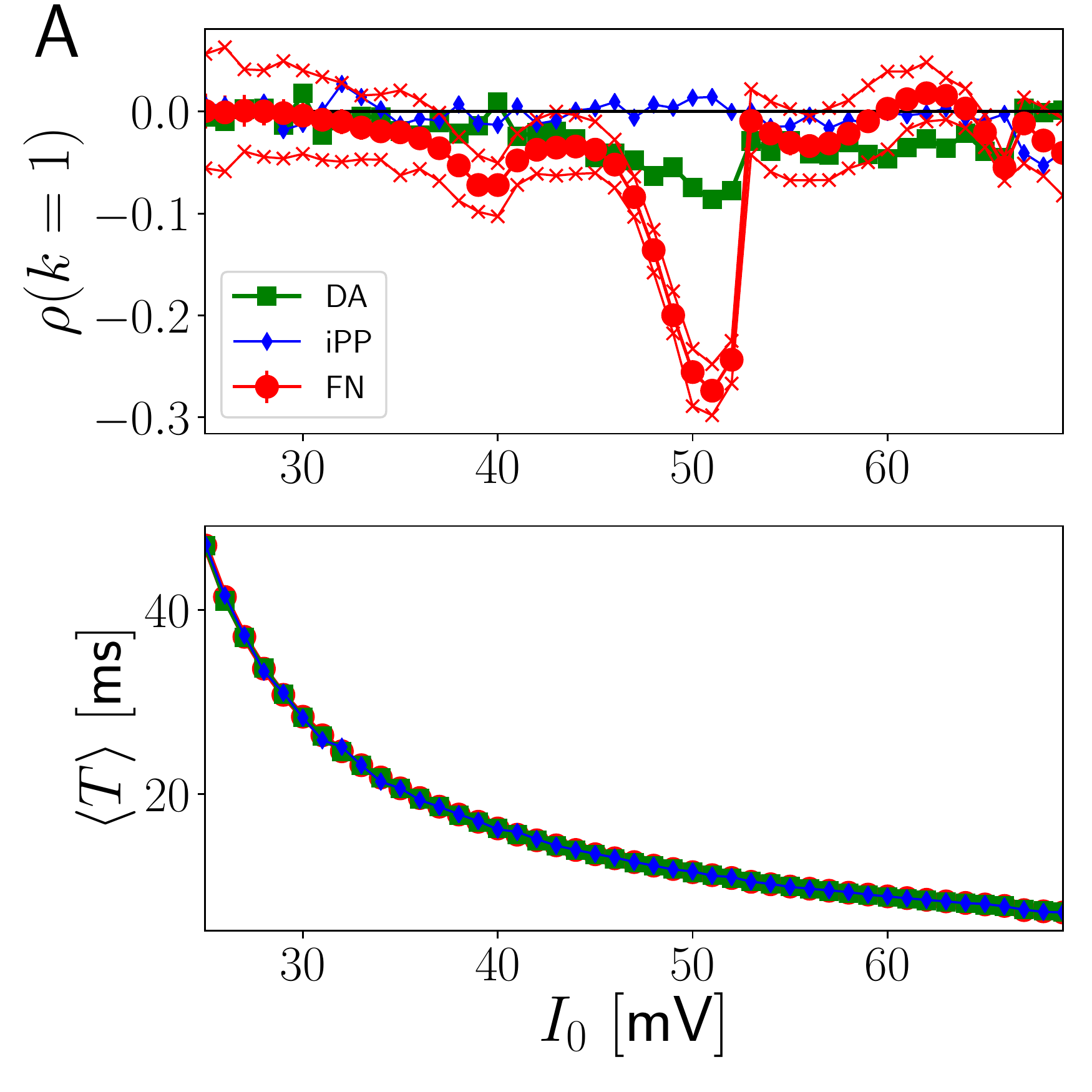}
\includegraphics[width = 0.4\textwidth]{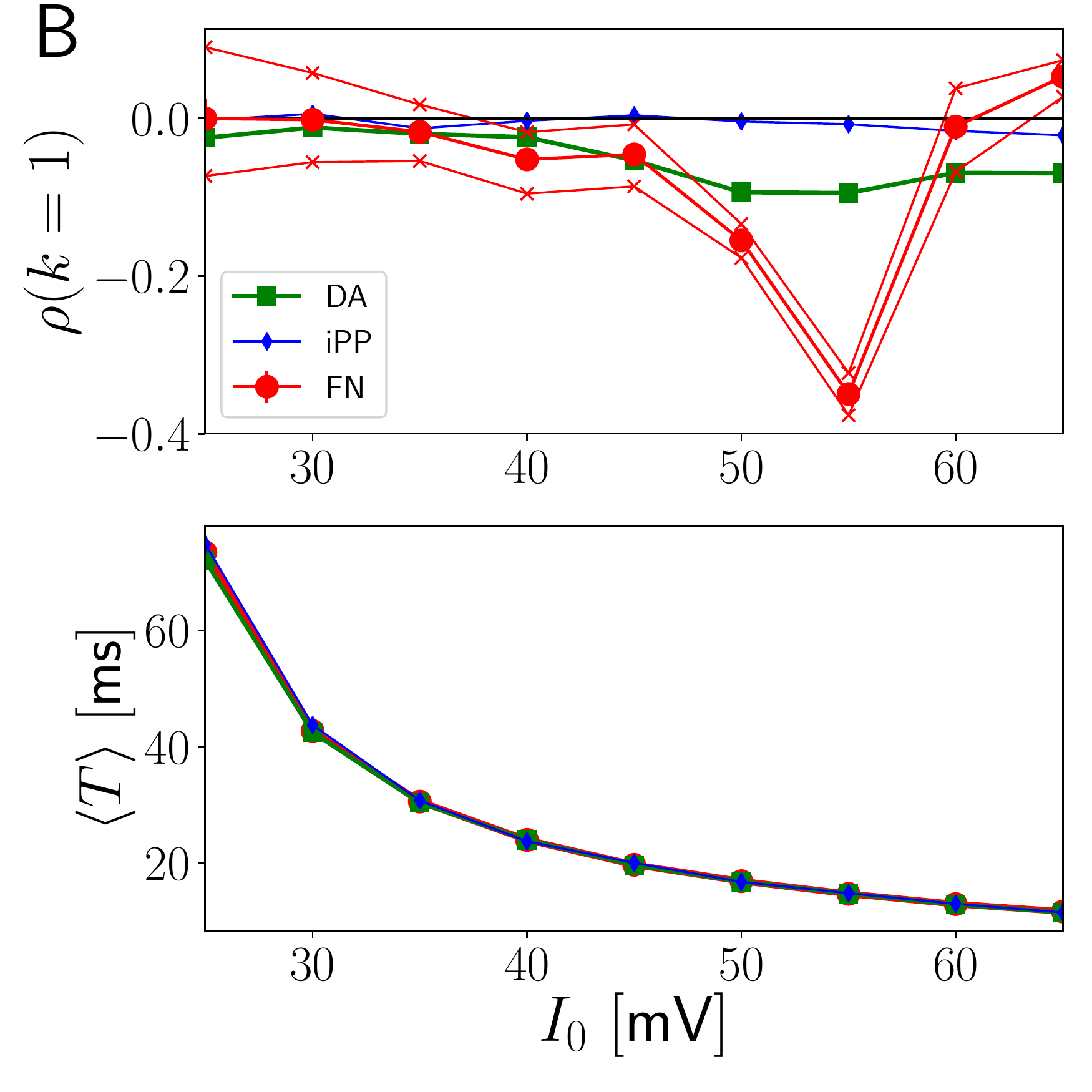}
\caption{Diffusion approximation for the mean network SCC (top of each panel) and ISI (bottom of each panel) as a function of $I_{0}$. \textbf{A}: $N = 500$. \textbf{B}: $N = 1000$. Red circles: full network simulation (FN). Green squares: single neuron offline simulation using the DA \eq \ref{eq:DA_alpha_J} with $\alpha_{J} = 1.0$. Errorbars for the full network simulation indicate $\pm$ one standard deviation of the SCC distribution. Parameter values: $C$-fixed scenario with $C = p(N-1)$. Simulation time $d = 200~\text{s}$.}
\label{fig_12}
\end{figure}

\begin{figure}[h!]
\includegraphics[width = 0.4\textwidth]{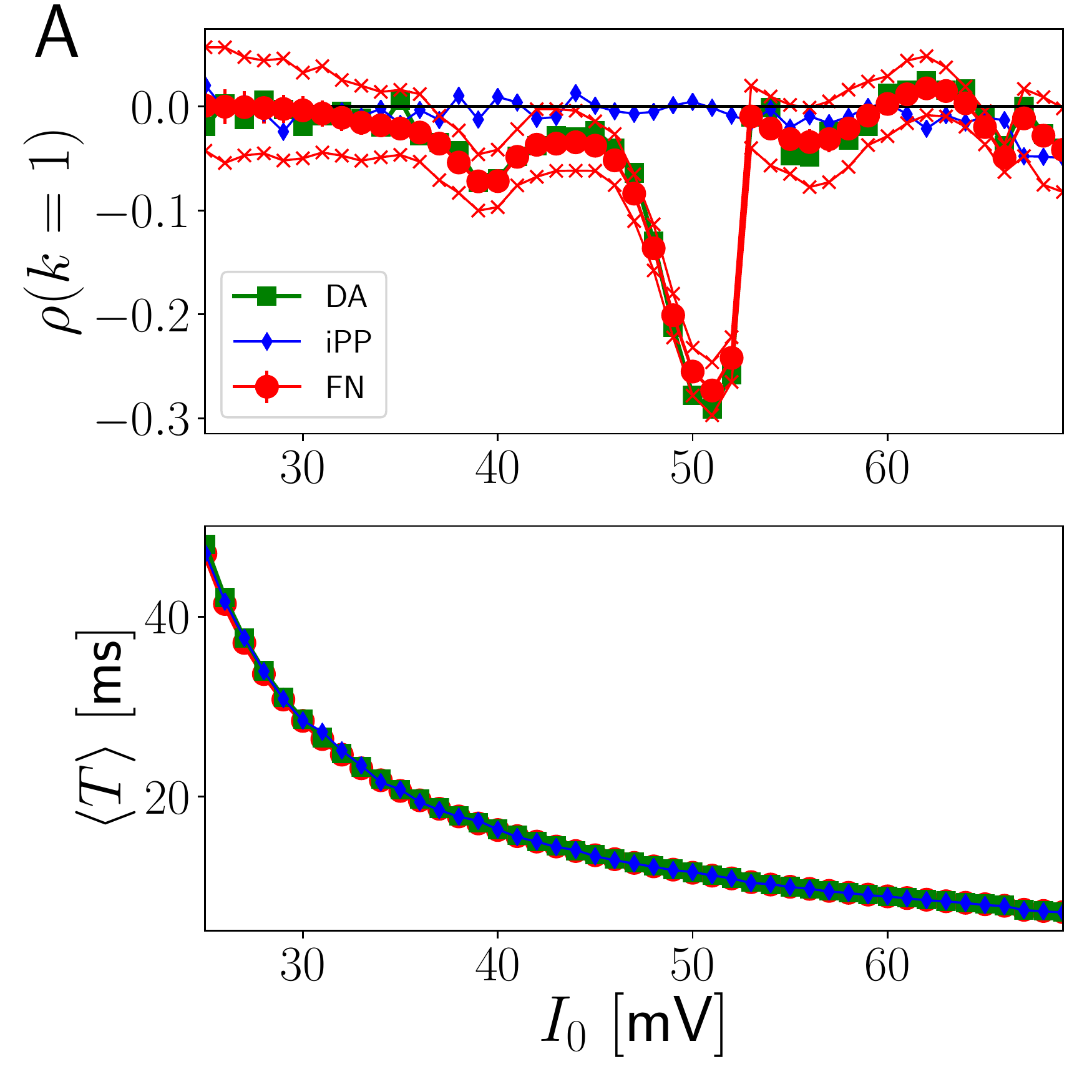}
\includegraphics[width = 0.4\textwidth]{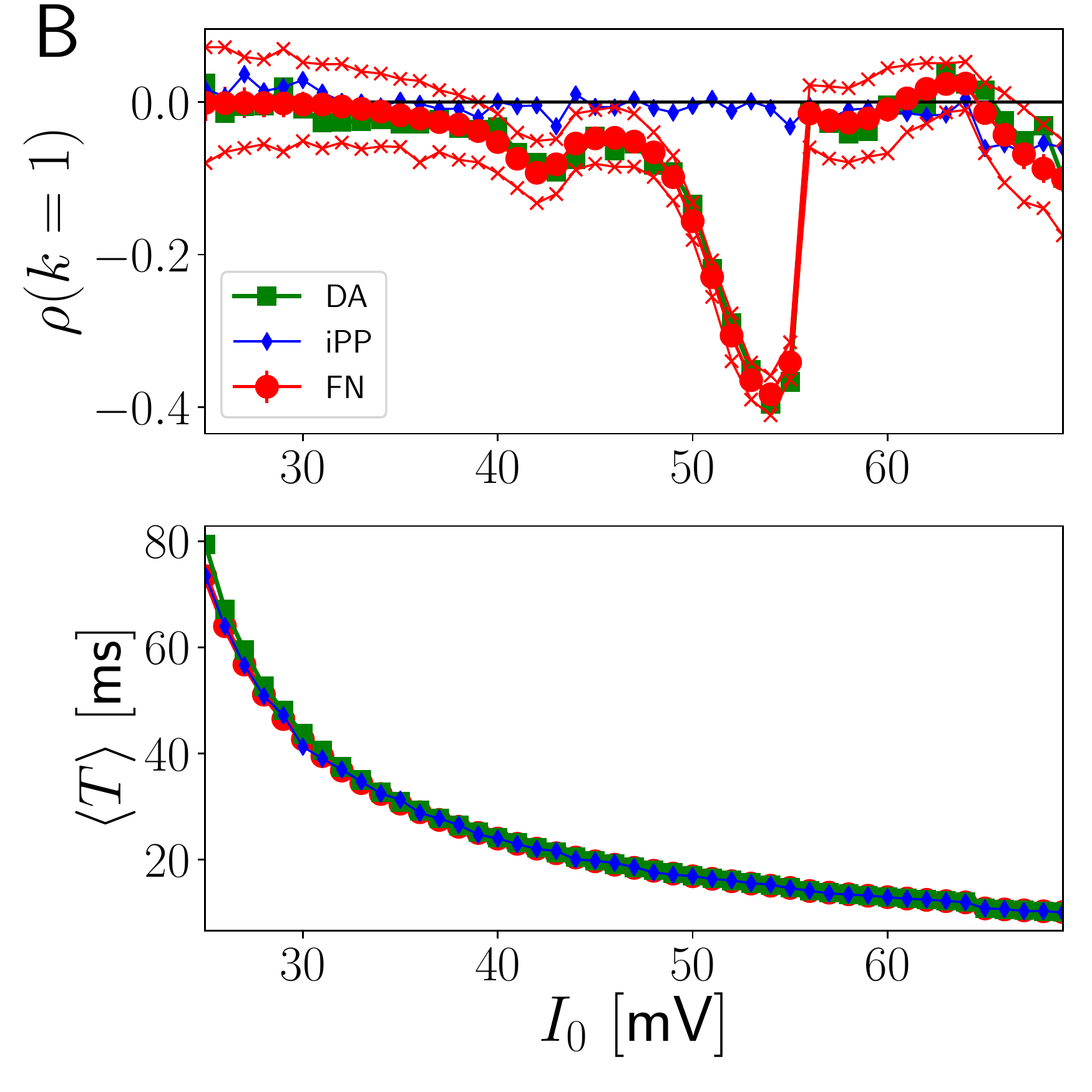}
\caption{Noise-reduced diffusion approximation for the mean network SCC (top) and ISI (bottom). \textbf{A}: $N = 500$. \textbf{B}: $N = 1000$. Red circles: full network simulation (FN). Blue diamonds: iPP approximation (iPP). Green squares: single neuron offline simulation using the DA \eq \ref{eq:DA_alpha_J} with $\alpha_{J} = 0$. Errorbars for the full network simulation indicate $\pm$ one standard deviation of the SCC. The thin red lines indicate the minimum and maximum of the online SCC distribution for each value of $I_{0}$. Parameter values as in \fig \ref{fig_12}.}
\label{fig_13}
\end{figure}

To understand the origin of the mismatch between the full network simulations and the DA at the onset of non-vanishing mean network ISI correlations, we computed the CVs of the spike train generated by the DA and compared it to the mean network CV (the CV averaged over all neurons in the network). We found that as $I_{0}$ exceeded approximately $40~\text{mV}$, the CV obtained by the DA was slightly larger than the mean network CV. Therefore, we decreased the contribution of the network activity to the noise in the DA by setting $\alpha_{J} = 0$ (see \eq \ref{eq:diffusion_approximation}). In \fig \ref{fig_13}, we show that this simple adjustment leads to good agreement between DA and full network simulations for both the mean network ISI and SCC.

\subsection{Effect of smoothing width}
\begin{figure}[h!]
\includegraphics[width = 0.4\textwidth]{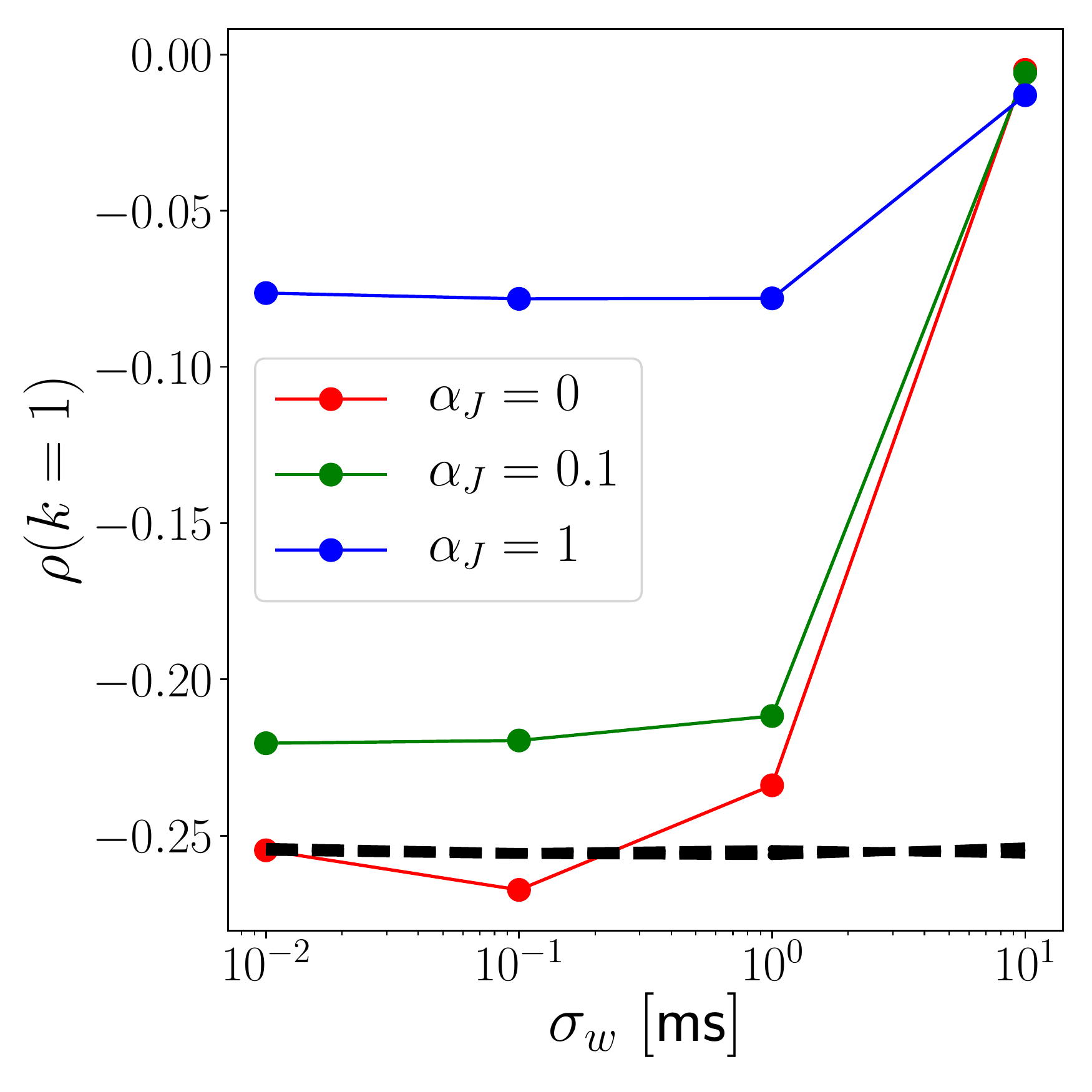}
\caption{Effect of smoothing width $\sigma_{w}$ and $\alpha_{J}$ on the performance of the DA (\eq \ref{eq:DA_alpha_J}). Solid lines: single neuron offline simulation using the DA \eq \ref{eq:DA_alpha_J} with three values of $\alpha_{J}$ as indicated in the legend. The black dashed lines are results from the online full network simulation, which do not depend on $\alpha_{J}$ or $\sigma_{w}$. Parameter values (analoguous to \fig \ref{fig_13} A for $I_{0} = 50~\text{mV}$): $N= 500, ~C = p(N-1)$ fixed, $p = 0.2$, $I_{0} = 50~\text{mV}$. Simulation length $d = 300~\text{s}$.}
\label{fig:fig_14}
\end{figure}

These results do not depend on the width of the smoothing kernel $\sigma_{w}$ as long as it is chosen small enough (\fig \ref{fig:fig_14}). Also, $\alpha_{J}$ must be chosen as small as possible for the noise-reduced DA to perform well.

\begin{figure}[h!]
\includegraphics[width = 0.4\textwidth]{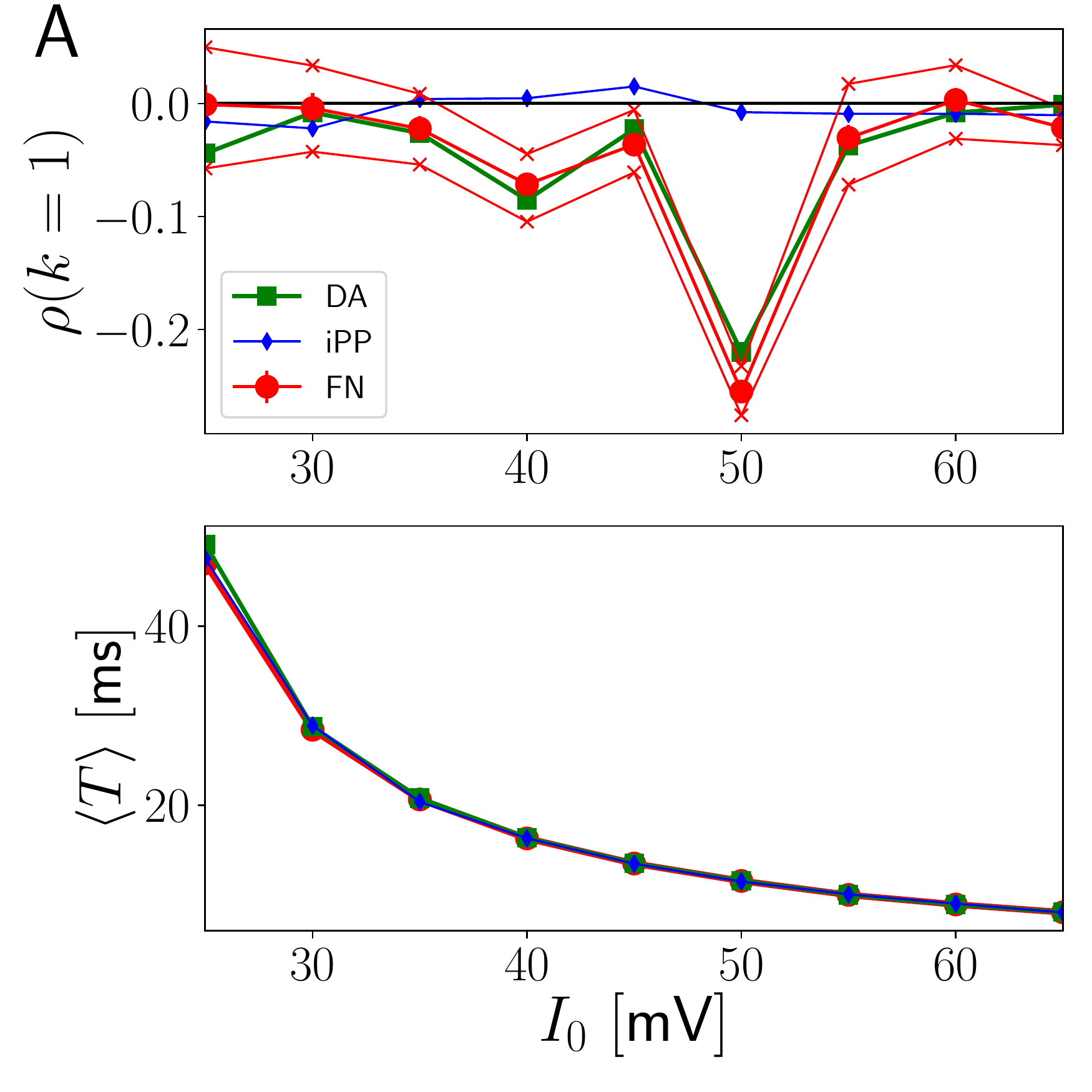}
\includegraphics[width = 0.4\textwidth]{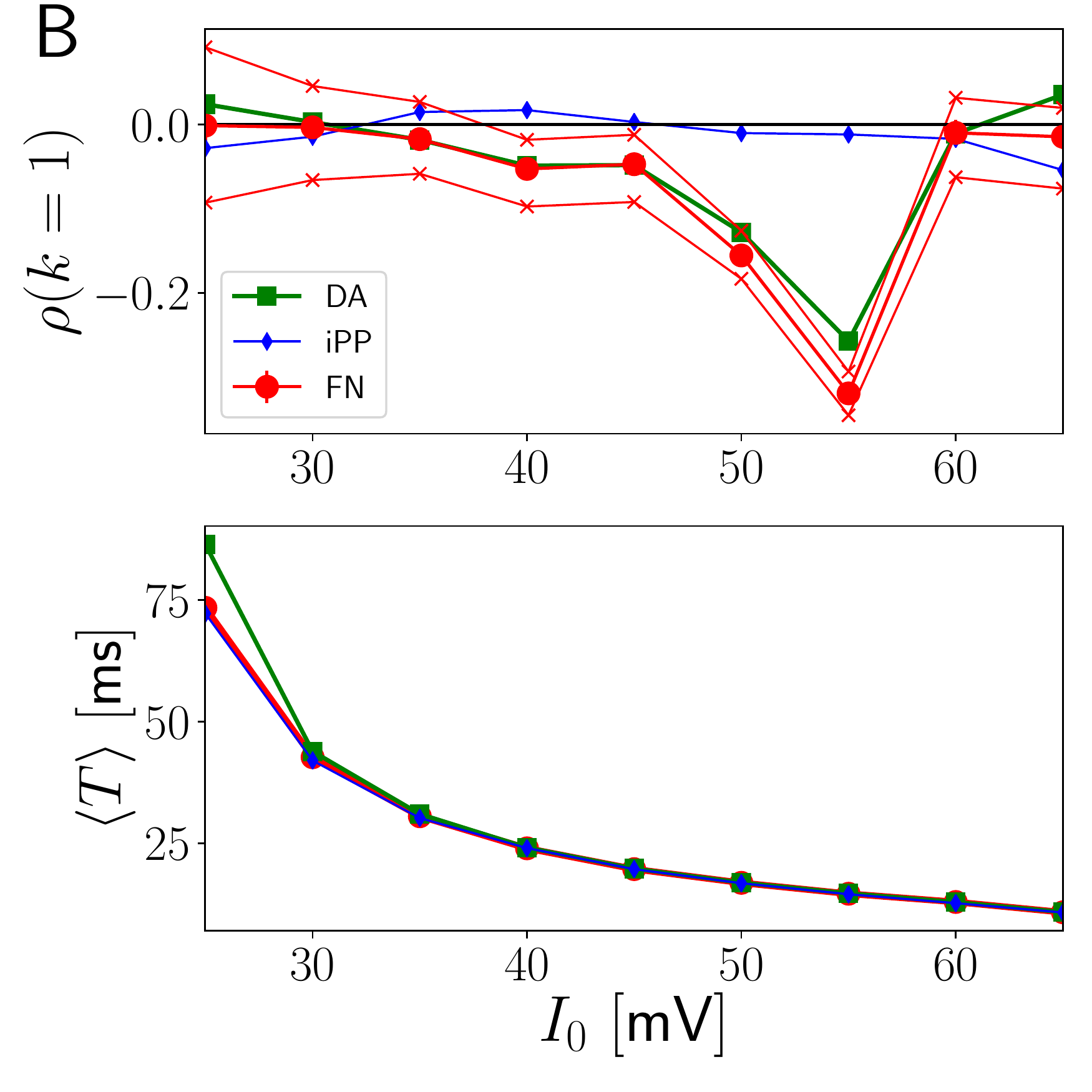}
\caption{Noise-reduced DA (\eq \ref{eq:DA_alpha_g} with $\alpha_{g} = 0.66$).
Diffusion approximation for the mean network SCC (top) and ISI (bottom). \textbf{A}: $N = 500$. \textbf{B}: $N = 1000$. Red circles: full network simulation (FN). Blue diamonds: iPP approximation (iPP). Green squares: single neuron offline simulation using the DA \eq \ref{eq:DA_alpha_g} with $\alpha_{g} = 0.66$. Errorbars for the full network simulation indicate $\pm$ one standard deviation of the SCC. The thin red lines indicate the minimum and maximum of the online SCC distribution for each value of $I_{0}$. Parameter values as in \fig \ref{fig_12}.}
\label{fig:fig_15}
\end{figure}

It is not necessary to scale down the noise in the particular way we have pursued so far with \eq \ref{eq:DA_alpha_J} to improve the accuracy of the DA. We show in \fig \ref{fig:fig_15} that a re-scaling of the noise term in the DA with a factor $\alpha_{g} = 0.66$ according to 

\begin{align}
\label{eq:DA_alpha_g}
\dot{X} = &\gamma_{t}\left(-X + I_{0} - CJ\nu(t-D) \tau_{t} \right) \\
 &+ \alpha_{g} \gamma_{t} \sqrt{ J^{2} C \nu(t-D)\tau_{t} + \sigma_{0}^{2}}\sqrt{\tau_{t}} \xi(t) \nonumber \, ,
\end{align}

has a similar effect on the accuracy of the DA as reducing the time-varying factor in the noise strength in \eq \ref{eq:DA_alpha_J}. However, this version of the DA performs slightly worse for the mean network ISI at small $I_{0}$, where it also underestimates the mean network CV (not shown). We found this particular value of $\alpha_{g}$ by systematic numerical simulations, gradually decreasing $\alpha_{g}$ from $1$ (equivalent to \eq \ref{eq:DA_alpha_J} with $\alpha_{J} = 1$). This confirms that the spiking dynamics of the full network generates input spike trains that are less stochastic than one would expect from a Poisson process with time-varying intensity.

\begin{figure}[!h]
\includegraphics[width = 0.4\textwidth]{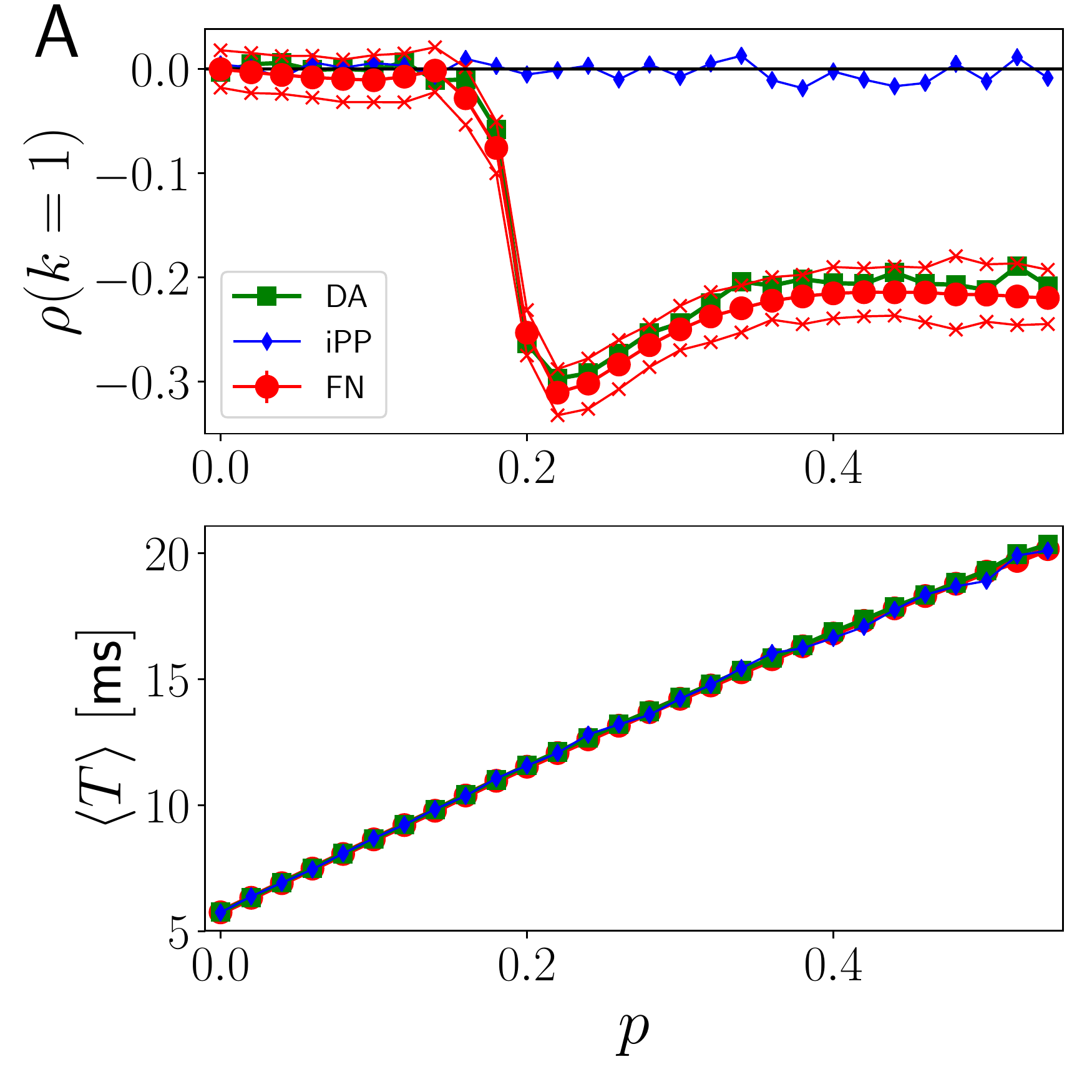}
\includegraphics[width = 0.4\textwidth]{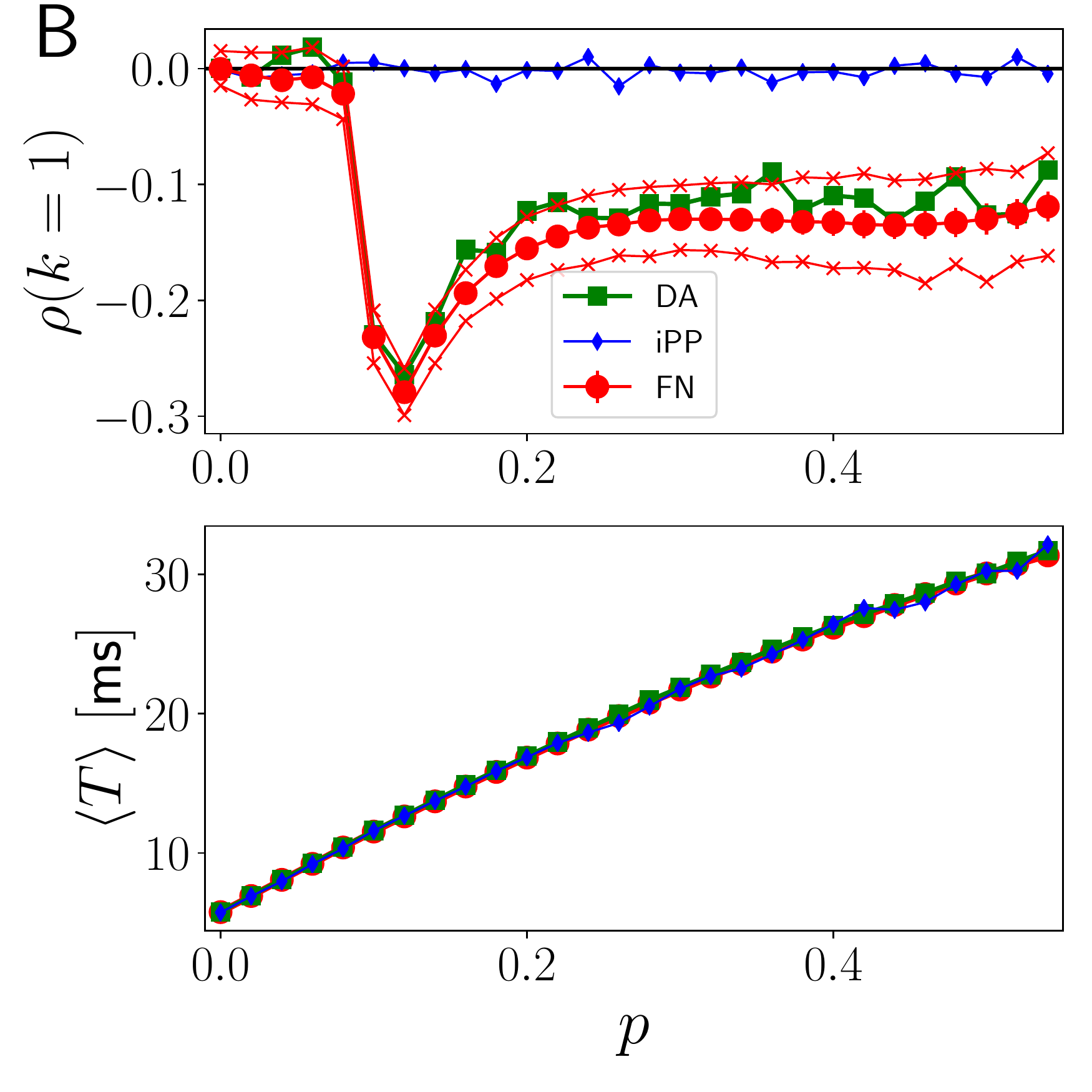}
\caption{DA for the mean network ISI and SCC is valid even for large $p$.
DA for the mean network SCC (top) and ISI (bottom). \textbf{A}: $N = 500$. \textbf{B}: $N = 1000$. Red circles: full network simulation (FN). Blue diamonds: iPP approximation (iPP). Green squares: single neuron offline simulation using the DA \eq \ref{eq:DA_alpha_J} with $\alpha_{J} = 0$. Error bars for the full network simulation indicate $\pm$ one standard deviation of the SCC. Parameter values: $C = p(N-1)$ fixed, $I_{0} = 50~\text{mV}$. Simulation time $d = 300~\text{s}$.}
\label{fig:fig_16}
\end{figure}

Finally, we show in \fig \ref{fig:fig_16} that the noise-reduced DA remains valid for a larger range of connection probabilities $p$, where we also observe a transition to negative mean network SCCs. Importantly, our standard value of $p=0.2$ is in the regime where the DA accurately reproduces the mean network ISI. Increasing $p$ for fixed $N$ is, in the $C$-fixed scenario, equivalent to an increase of $N$ for fixed $p$, for which we have already shown that the mean network SCC decreases to negative values for large enough $I_{0}$ (black horizontal dashed line in \fig \ref{fig_2}).

In summary, our explorations of three forms of the DA (two of which are effectively noise-reduced) using offline simulations show that it is possible to obtain conditions where a DA with the proper smoothed spike input $A(t)$ produces correct first and second order ISI statistics. This will help constrain the search for a self-consistent effective neuron model that can produce such statistics.

\section{Discussion}
\label{sec:discussion}
We have studied ISI correlations in spike trains of neurons embedded in purely inhibitory networks with both current- and conductance-based synapses. Our results reveal that these simple networks in general do not generate renewal dynamics in their spike trains. This is not surprising given their propensity to oscillate (\fig \ref{fig_1}). However, the manner in which serial ISI correlations appear and their non-monotonic behavior (which implies maximal ISI correlation at intermediate parameter values) and link to SCC variance in the $P$-fixed case (\fig \ref{fig_4}) are new features revealed by our analysis. 

In particular, strong negative serial correlations are observed at an intermediate value for the bias drive and network size in networks without and with a refractory period (\figs \ref{fig_2} - \ref{fig_6}). We have also shown that the mean network SCC is maximally negative at an intermediate value of the network oscillation strength (as quantified by the peak value of the PSD of the network activity $A(t)$, \fig \ref{fig:coherence_frequency}) as it decreases below zero with increasing bias current or system size.

In conductance-based networks, negative ISI correlations also arise generically with increasing bias drive for both slow and fast gamma network oscillations (\figs \ref{fig_7} - \ref{fig_9}). 

Importantly, our results highlight the importance of nearly all input spikes for shaping the mean network SCC (\fig \ref{fig_10}). Even if the input statistics to a given neuron are non-Poissonian (\fig \ref{fig_11}), the onset of negative temporal correlations can be approximately quantified by an effectively noise-reduced diffusion approximation (\eq \ref{eq:DA_alpha_J}), showing that the network dynamically suppresses noise at increasing values of the bias current drive. Whereas this decrease of the noise intensity does not have a strong effect on the validity of the DA for the mean network ISI, we have shown that it has a pronounced effect for the mean network SCC (\figs \ref{fig_12} - \ref{fig:fig_16}). 

Overall, our results constitute a first step to understand how pre-synaptic spiking shapes the statistics of post-synaptic spiking in neural networks (\figs \ref{fig_10}- \ref{fig_11}), a question that has recently received increased interest in the biological literature, with positive ISI correlations observed in the zebrafish lateral line system \cite{song_et_al_2018} and negative correlations observed in the auditory system afferent neurons \cite{peterson_et_al_2014}. In these systems, it is thought that positive ISI correlations arise from slowly drifting firing rates caused by heterogeneous innervation of more than one haircell synapse, whereas negative ISI correlations arise from synaptic depletion alone. It would be interesting to extend our results to these two scenarios, which would require the inclusion of non-stationary rate dynamics and synaptic plasticity into our model.

We would also expect transitions to negative mean network SCCs in networks of Hodgkin-Huxley or IF neurons with adaptation, as long as noisy oscillatory states are present. This is because the detailed voltage dynamics are only important insofar as they interact with the global rhythm, i.e. the form of the subthreshold voltage time course is not crucial for the onset of negative mean network SCCs \cite{geisler_brunel_wang_2005}.

Concerning a possible full analytical understanding of SCC transitions, the existence of ISI correlations points to stochastic dynamics in dimension greater than one, and an analytic explanation of our results is beyond the scope of our study. Recent first-passage time results for a two-dimensional Wiener process \cite{sacerdote_tamborrino_zucca_2016} nevertheless give hope that the present numerical approach will eventually lead to an analytical description.

It is also an interesting subject for future study to determine when non-vanishing ISI correlations exist also in networks with both excitatory and inhibitory subpopulations. Furthermore, how our results relate to transitions in networks with these two types of neurons observed by increasing coupling strength \cite{ostojic_2014, wieland_bernardi_schwalger_lindner_2015} and whether they can be computed using the framework of \cite{schwalger_droste_lindner_2015} or the framework of doubly stochastic point processes \cite{lowen_teich_1992} remain open questions.

\section{Acknowledgments}
We would like to thank Raoul-Martin Memmesheimer, Sven Goedeke, Alexandre Payeur and Richard Naud for interesting discussions and helpful comments. We would like to thank NSERC Canada (grant number 121891-2013) for funding this work and the German Federal Ministry of Education and Research (BMBF) for support via the Bernstein Network (Bernstein Award 2014, 01GQ1501 and 01GQ1710). W.B. thanks Denisa Dervishi and Karl Braun for hospitality.

\appendix
\section{Details for network simulation}
\subsection{Simulation details}
\label{sec:simulation_details}
We implemented all simulations in \texttt{brian2} \cite{brian2}.  The integration time step was set to $h = 10^{-2}~\text{ms}$ if not mentioned otherwise and a standard Euler-Maruyama scheme was used. Except in \fig \ref{fig_6}, the refractory period is set to $0~\text{ms}$ for the current-based synapse model.
To ensure that we are in a stationary regime, we discard the first $1000$ ISIs both for the computation of the mean network SCC and ISI. The statistics of the SCC and ISI are computed across long realizations (at least $100~\text{s}$) of the network activity. Increasing the duration did not change the results presented here. Note that the network activity of interest will sometimes consist in a rhythm, in which case the ISI distribution, and the SCC, vary across the duration of the period of the rhythm. In this case, the averages are taken over many cycles of the rhythm to obtain a mean and standard deviation of the ISI and the SCC during the rhythm.  

\subsection{Connectivity details}
\label{sec:connectivity_details}
For the $P$-fixed case, we use the standard syntax of \texttt{brian2} in the \texttt{Synapses.connect()} object.
We wire up $C$-fixed networks by first making a masterlist of all the neurons, each labeled by its number between $1$ and $N$. To choose which neurons are connected to e.g. the neuron with label $k$, this neuron $k$ is deleted from the list. Then the remaining list is randomly permuted. Finally, the first $C$ neurons from that list are chosen as the pre-synaptic neurons to neuron $k$. The procedure is repeated for every neuron in the masterlist.

\subsection{Spectral measures}
\label{sec:spectral_measures}
A smoothed version $A(t)$ of the population activity is computed using the function \texttt{smooth\char`_rate()} of the \texttt{PopulationRateMonitor} class in \texttt{brian2}. This is the instantaneous firing rate of the network. If not mentioned otherwise, $A(t)$ is obtained from the pooled spike trains of all neurons by convolution with a Gaussian rectangular or Gaussian kernel of width $\sigma_{w}$. For illustration purposes, we choose $\sigma_{w} = 1~\text{ms}$; for computations involving the diffusion approximation, we choose a smaller $\sigma_{w} = 0.01~\text{ms}$ or $\sigma_{w} = 0.1~\text{ms}$. We stress that the choice of $\sigma_{w}$ does not influence our results as long as it is small enough to fully capture the dynamics of the population activity $A(t)$, i.e. it is important not to smoothen out too much the fluctuations present in $A(t)$. 

For the computation of spectra (both the PSD of the network activity as well as the averaged single-neuron spectrum), we used two functions provided in \cite{neuronal_dynamics_book}. For the average single-neuron spectrum, only $100$ neurons were included in the average. For both types of spectra, the mean activity before taking the Fourier transform was subtracted and the first $1000~\text{ms}$ of the simulation period were discarded to obtain stationary time series. The code can be found in the \texttt{github} repository \url{https://github.com/EPFL-LCN/neuronaldynamics-exercises}.

For the PSD of the network activity, the frequency resolution was set to $2~\text{Hz}$ and the time series for $\nu$ was split into $6$ parts that were Fourier-transformed independently and then averaged to obtain the PSD. Similarly, the averaged single-neuron spectrum $\langle S_{xx}(f) \rangle$ was computed with a sampling rate of $6000~\text{Hz}$. It was checked that the functions \texttt{scipy.signal.periodogram} and \texttt{scipy.signal.welch} gave the same results for both types of spectra.

\section{SCC for a noisy periodic oscillation}
\label{sec:appendix_1}
Consider firing times $t_{i}, i \geq 1$ generated according to

\begin{equation}
 t_{i} = n_{i} \langle t \rangle + \xi_{i} \, ~ \langle \xi_{i} \rangle = 0, \langle \xi_{i} \xi_{j} \rangle = \sigma^{2} \delta_{ij} \, ,
\end{equation}

where $n_{i} = 1, 2, 3,...$ and $\langle t \rangle$ is the period of firing around which the spike times are perturbed. The $i$th ISI is given by

\begin{equation}
 T_{i} = t_{i} - t_{i-1} \, .
\end{equation}

It follows that $\langle T_{i}\rangle = \langle t \rangle$. Similarly, we can show that $\mathbb E \left(T_{i+1} T_{i} \right) = \langle t \rangle^{2} - \sigma^{2}$. Moreover, we have

\begin{equation}
\text{Var}(T_{i}) = \langle T_{i}^{2} \rangle - \langle T_{i} \rangle^{2} = \langle t \rangle^{2} + 2 \sigma^{2} - \langle t \rangle^{2} \, .
\end{equation}

Hence, the SCC is given by

\begin{equation}
\rho(k=1) = \frac{\langle t \rangle^{2} - \sigma^{2} - \langle t \rangle^{2} }{\langle t \rangle^{2} + 2\sigma^{2} - \langle t \rangle^{2}} = -\frac{1}{2} \, .
\end{equation}

This result can be generalized to the case when $n_{i}$ itself is a random number drawn from a certain (e.g. binomial) distribution.

\section{Inhibitory integrate-and-fire neuron model with conductance-based synapses}
\label{sec:coba}

We consider a model with parameters similar to \cite{donoso_et_al_2018}. The parameters are given by $C_{I} = 100~\text{pF}$, $g^{I}_{l} = 10~\text{nS}$ (so that the membrane time constant is $10~\text{ms}$), $E^{I}_{\text{rest}} = -65~\text{mV}$, $E^{I}_{\text{inh}} = -75~\text{mV}$. When $X_{i}(t)$ exceeds $X^{I}_{\text{thresh}} = -52~\text{mV}$, it is reset to $X^{I}_{\text{reset}} =  -67~\text{mV}$, followed by a refractory period of duration $\tau^{I}_{\text{ref}} = 1~\text{ms}$. 
In the absence of synaptic inputs, the rheobase is $I_{0} = 0.13~\text{nA}$. In our simulations, we choose $I_{0} = 0$. The only excitatory input to the population is delivered via excitatory Poisson processes. Each neuron receives input from  $n^{\text{P}}_{I}$ independent Poisson processes, each with frequency $f^{\text{P}}_{I}$. Each of these random inputs induces instantaneous jumps of the membrane potential $X_{i}$ by an amplitude $\kappa^{\text{P}}_{I} = 0.1~\text{mV}$.


The latency between a presynaptic spike and the start of the increase of $g^{I}_{\text{inh}}$ is $\tau_{l} = 1~\text{ms}$. The remaining parameters are given by $g^{I}_{\text{inh}, \text{peak}} = 5.0~\text{nS}$, $\tau^{I}_{\text{inh}, d} = 1.2~\text{ms}$ and $\tau^{I}_{\text{inh}, r} = 0.45~\text{ms}$.

\bibliography{literature_late_2017}

\end{document}